\documentclass[aps,prl,showpacs,floatfix,twocolumn,superscriptaddress]{revtex4}
\usepackage{graphicx,amsmath,amssymb,dcolumn}

\begin{document}

\title{Lattice QCD study of a five-quark hadronic molecule}
\author{M.S. Cook}
\affiliation{Physics Department, FIU-University Park, Miami, Florida 33199}
\affiliation{Department of Physics, Converse College, Spartanburg, South Carolina 29302}
\author{H.R. Fiebig}
\affiliation{Physics Department, FIU-University Park, Miami, Florida 33199}

\date{\today}

\begin{abstract}
We compute the ground-state energies of a
heavy-light $K$--$\Lambda$ like system as a function of the relative
distance $r$ of the hadrons. The heavy quarks, one in each hadron, are
treated as static. Then, the energies give rise to an adiabatic
potential $V_a(r)$ which we use to study the structure
of the five-quark system.
The simulation is based on an anisotropic and asymmetric lattice
with Wilson fermions. Energies are extracted from spectral density
functions obtained with the maximum entropy method.
Our results are meant to give qualitative insight: Using the resulting adiabatic
potential in a Schr\"odinger equation produces bound state wave functions
which indicate that the ground state of the five-quark system resembles a hadronic
molecule, whereas the first excited state, having a very small rms radius,
is probably better described as a five-quark cluster, or a pentaquark.
We hypothesize that an all light-quark pentaquark may not exist, but in the
heavy-quark sector it might, albeit only as an excited state.
\end{abstract}


\maketitle

\section{Introduction}

Lattice QCD studies of hadron-hadron interactions are the gateway to
nuclear physics through first principles \cite{Fiebig:2002kg}. 
From a lattice simulation point of view the nucleon-nucleon interaction is
unquestionably the most challenging case \cite{Savage:2005}, and it might
not be resolved in the foreseeable future.
However, interactions in other two-hadron systems are worth
investigating as well, because new insights into the structural features of
already discovered \cite{Yao:2006px} or yet unknown baryon resonances
may emerge. In particular, one may ask if some of those may be understood
as hadronic molecules, like the deuteron, or if more compact 
clusters, like a pentaquark \cite{Diakonov:2003jj}, may also exist.

We are here interested in pairs of hadrons containing one
heavy quark each. In such systems a relative, residual, interaction is
a well-defined concept.
In the spirit of the Born-Oppenheimer approximation, the (slow) heavy quarks
serve to define the centers of the two hadrons while the (fast) light quarks and gluons
provide the physics of the interaction.
Exploratory studies along those lines have been done in the context of meson-meson and
baryon-baryon systems \cite{Arndt:2003vx,Michael:1999nq,Mihaly:1997ue}.

\begin{figure}[h]
\includegraphics[angle=0,width=55mm]{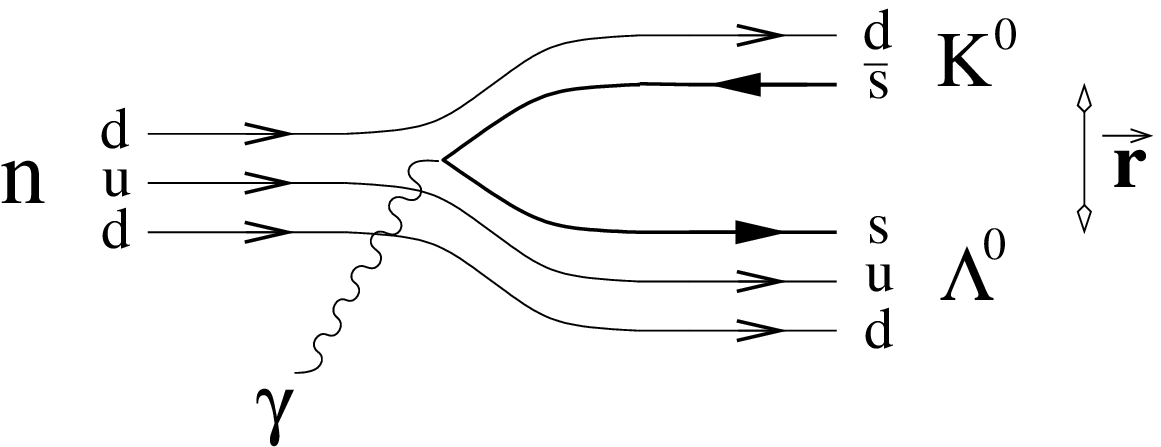}\\
\includegraphics[angle=0,width=55mm]{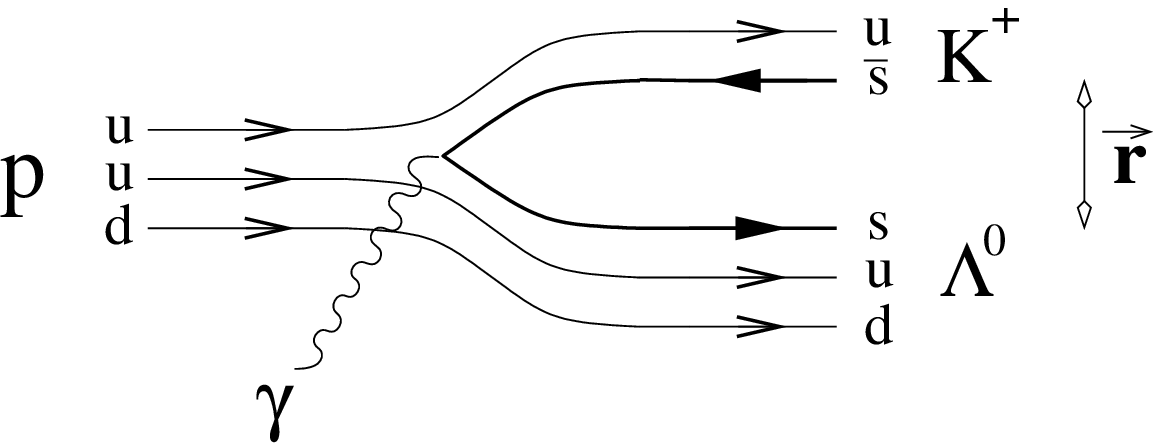}
\caption{Schematic view of $K$--$\Lambda$ like molecule production and
illustration of the current scheme for extracting a hadronic interaction
at relative distance $\vec{r}$.
Thick lines indicate heavy-quark propagators, and thin
lines depict light quark propagators.}
\label{fig:KLproduction}
\end{figure}

Specifically, we here investigate a heavy-light meson-baryon (five-quark) hadron
with the quantum numbers of an S-wave $K$--$\Lambda$ system.
The heavy-quark propagator is treated in the static approximation.
This allows us to compute the total energy as a function of the relative
distance $r$ between the heavy quarks, viz. hadrons.
A production scheme is illustrated in Fig.~\ref{fig:KLproduction}.
The resulting adiabatic (Born-Oppenheimer) potential
$V_a(r)$ then may be used to address the possibility of molecule-like,
or other, structures.

In the static approximation to the heavy-quark propagator
the potential $V_a(r)$ extracted from the lattice equally applies
to the systems $K$--$\Lambda$, $D$--$\Lambda_c$, and $B$--$\Lambda_b$,
the $s$-flavored quarks being replaced with $c$ and $b$ flavors, respectively.
Within the framework of a Schr{\"o}dinger equation those systems can
be studied using their respective physical reduced masses, provided of course,
one is willing to accept the limitations of the adiabatic approximation
and the non-relativistic nature of the framework.
This is the main reason for classifying  our results as qualitative. 

\section{Operator construction}

\begin{figure}[h]
\noindent\rule{1mm}{0mm}
\includegraphics[angle=0,width=27mm]{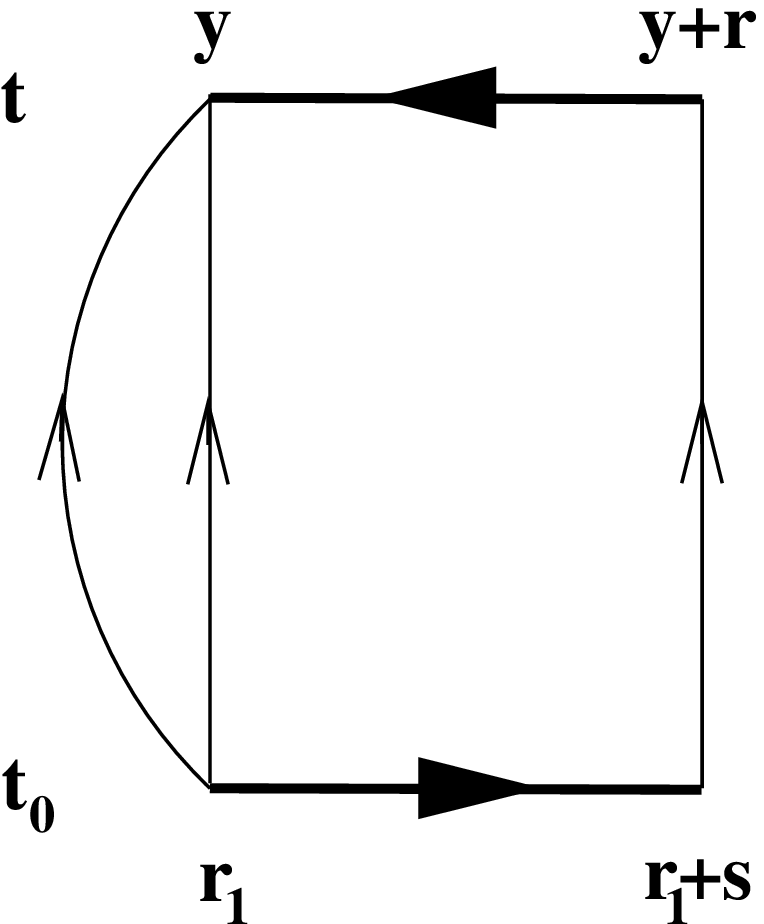}\hspace{12mm}
\includegraphics[angle=0,width=27mm]{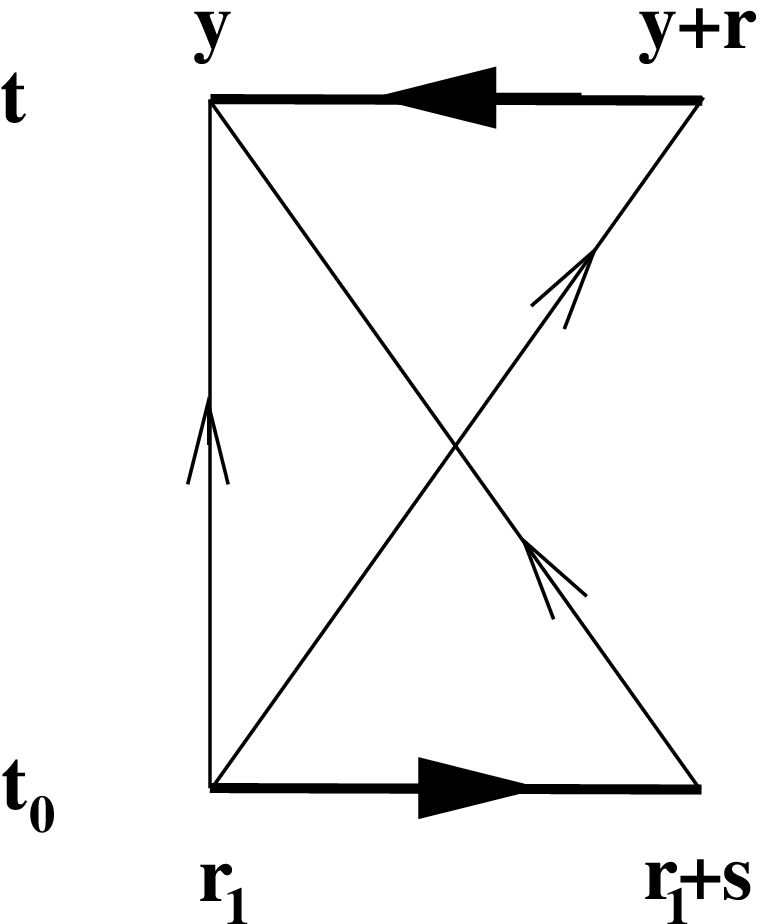}\vspace{4mm}\\
\hspace{1mm}
\includegraphics[angle=0,width=27mm]{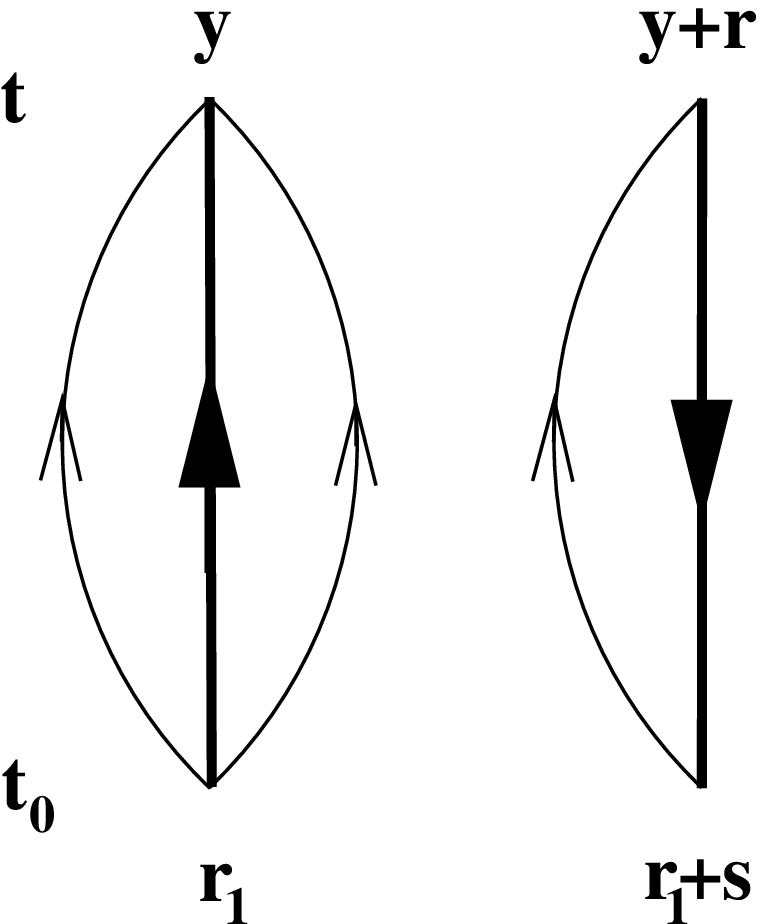}\hspace{12mm}
\includegraphics[angle=0,width=27mm]{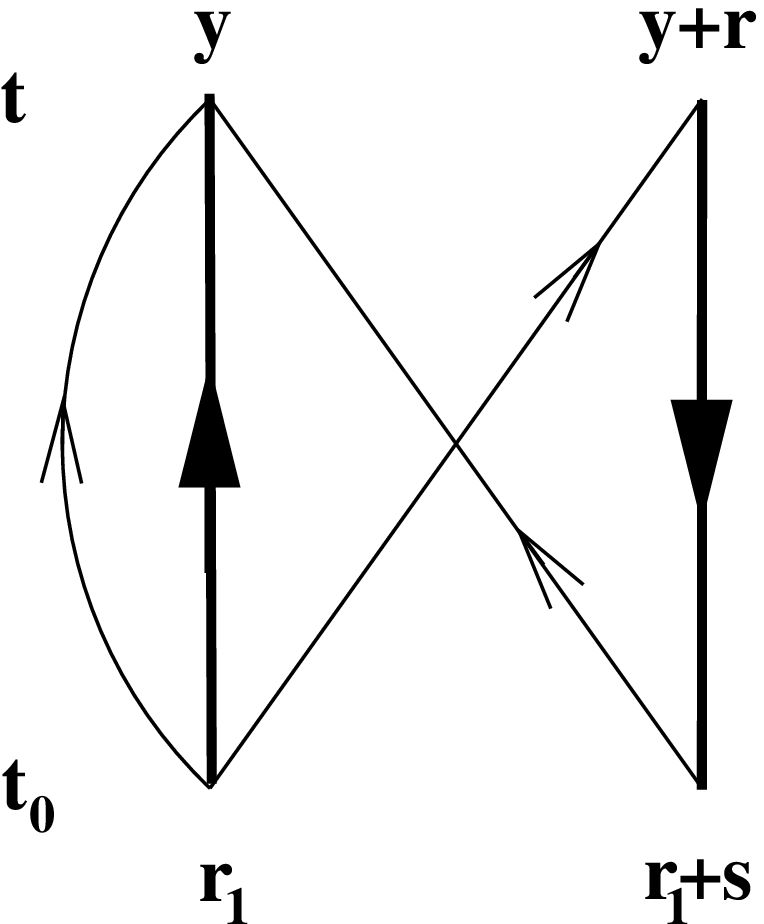}
\caption{Diagrammatic representation of (\protect\ref{CHG}).
Thick lines indicate heavy-quark propagators, and thin
lines depict light (u and d) quark propagators.}
\label{fig:CHG}
\end{figure}

The lattice simulation requires two-hadron interpolating fields. Those are
constructed using as building blocks standard local operators for the $K^+$
and $\Lambda^0$ particles \cite{Mon94}.
They are placed at relative distance $\vec{r}$ and then projected to total
momentum zero
\begin{equation}\label{Op2}
{\cal O}_\alpha(\vec{r};t)=N^{-1/2}\sum_{\vec{x}}\sum_{\vec{y}}
\delta_{\textstyle\vec{r},\vec{x}-\vec{y}}\,
K^+(\vec{x}t)\Lambda^0_\alpha(\vec{y}t)\,.
\end{equation}
Here the normalization $N$ is the spatial lattice volume, the sums are over lattice sites,
and $\alpha$ is a Dirac spinor index. Then, with
\begin{equation}
\bar{\cal O}_\mu(\vec{r};t)={\cal O}^\dagger_\alpha(\vec{r};t)\gamma_{4,\alpha\mu}
\end{equation}
the time correlation function
\begin{equation}\label{corfun}
C(t,t_0)= \langle {\cal O}_\mu(\vec{r};t) \bar{\cal O}_\mu(\vec{s};t_0) \rangle
-\langle {\cal O}_\mu(\vec{r};t) \rangle\langle \bar{\cal O}_\mu(\vec{s};t_0) \rangle\,,
\end{equation}
where $\vec{s}$ is the relative distance at the source,
can be expressed in terms of fermion propagators. The flavor assignment
$K^+\Lambda^0\sim \bar{s}u\,uds$ causes the separable term in (\ref{corfun}) to vanish.
Writing $H(\vec{x}t,\vec{y}t_0)$ and $G(\vec{x}t,\vec{y}t_0)$
for the heavy (s) and light (u,d) quark propagators, respectively, one obtains
\begin{eqnarray}\label{CHG}
C(t,t_0)&=&\langle\;{\textstyle\sum_{\vec{y}}}\;
[H(\vec{y}t,\vec{y}+\vec{r}t)H(\vec{r}_1+\vec{s}t_0,\vec{r}_1t_0)\nonumber\\
&&-H(\vec{y}t,\vec{r}_1t_0)H(\vec{r}_1+\vec{s}t_0,\vec{y}+\vec{r}t)]\times\nonumber\\
&&\phantom{-}G(\vec{y}t,\vec{r}_1t_0)
[G(\vec{y}t,\vec{r}_1t_0)G(\vec{y}+\vec{r}t,\vec{r}_1+\vec{s}t_0)\nonumber\\
&&-G(\vec{y}t,\vec{r}_1+\vec{s}t_0)G(\vec{y}+\vec{r}t,\vec{r}_1t_0)]\;\rangle\,.
\end{eqnarray}
For clarity the rather involved color and spin index structure is not shown in (\ref{CHG}).
Also, translational invariance of the gauge field average $\langle\,\rangle$
has been used to arrive at the
above expression, and utilizing this freedom, an
arbitrary space site $\vec{r}_1$ was introduced to fix the source locations.
A diagrammatic representation of (\ref{CHG}) is shown in Fig.~\ref{fig:CHG}.

The heavy-quark propagators are employed in the static approximation.
For (unimproved) Wilson fermions with hopping parameter $\kappa$ this
means that the propagator is taken in the limit $\kappa\rightarrow0$ ,
resulting in
\begin{equation}\label{Hstatic}
H(\vec{x}t,\vec{y}t_0) = \delta_{\textstyle \vec{x},\vec{y}}\;(2\kappa)^{t-t_0}
\frac{1}{2}(1+\gamma_4)\,\mathcal{U}^\dagger(\vec{x};t_0t)\,,
\end{equation}
where $\mathcal{U}(\vec{x};t_0t)$ is the product of $SU(3)$ link variables
along a straight line from $(\vec{x}t_0)$ to $(\vec{x}t)$ \cite{Mon94}. 
The factor $(2\kappa)^{t-t_0}$
gives rise to only a constant global energy shift $\Delta m=-\ln(2\kappa)$,
which we will ignore.

The distance $\vec{r}=0$ is rather special \cite{Michael:1999nq} because a color singlet
operator, as realized by (\ref{Op2}), can also be achieved by a ``color twisted'' version
of (\ref{Op2}) where quarks across the hadrons $K^+$ and $\Lambda^0$,
now at the same location, are combined into a color singlet.
Because we do not consider color twisted operators in this work, we restrict ourselves
to non-zero relative distance\footnote{Experience
has shown \protect\cite{Mihaly:1997ue} that zero-distance operators give rise to
discontinuous, non-smooth, behavior of the potential at $\vec{r}=0$.}.
Thus, using $H(\vec{y}t,\vec{y}+\vec{r}t)\propto \delta_{\vec{r},\vec{0}}$
and $H(\vec{r}_1+\vec{s}t_0,\vec{r}_1t_0)\propto \delta_{\vec{s},\vec{0}}$,
the first two terms in (\ref{CHG}), and accordingly
the top two diagrams in Fig.~\ref{fig:CHG} vanish, and only
the last two of those make a contribution to
the correlation function for non-zero relative distance.
By way of (\ref{Hstatic}) those contributions are proportional to
$\delta_{\vec{y},\vec{r}_1}\delta_{\vec{r},\vec{s}}$.
Introducing $\vec{r}_2=\vec{r}_1+\vec{s}$, to replace $\vec{s}$, the
correlation function (\ref{CHG}) becomes
\begin{eqnarray}\label{C}
C(t,t_0)&=&\delta_{\textstyle\vec{r},\vec{r}_2-\vec{r}_1}
\;\langle H(\vec{r}_1t,\vec{r}_1t_0)H(\vec{r}_2t_0,\vec{r}_2t)\\
&&G(\vec{r}_1t,\vec{r}_1t_0)
[-G(\vec{r}_1t,\vec{r}_1t_0)G(\vec{r}_2t,\vec{r}_2t_0)\nonumber\\
&&+G(\vec{r}_1t,\vec{r}_2t_0)G(\vec{r}_2t,\vec{r}_1t_0)]\;\rangle
\quad\mbox{for}\quad\vec{r}\neq 0\nonumber\,.
\end{eqnarray}
As a consequence of the static approximation,
the site sum $\sum_{\vec{y}}$ has vanished from (\ref{CHG}), and thus, is
unfortunately no longer working to improve statistics. 

The final correlator we use is extended from (\ref{C}) to a $K\times K$ matrix by
employing several levels $k=1\ldots K$ of operator smearing. The procedure
amounts to replacing in (\ref{Op2}) all light-quark fields $\psi,\bar{\psi}$
with smeared fields $\psi^{\{k\}},\bar{\psi}^{\{k\}}$.
We have used Wuppertal-style fermion smearing \cite{Alexandrou:1994ti}
along with APE-style gauge field fuzzing \cite{Alb87a}.
The implementation specific to the asymmetric lattice used here
is discussed below.
No smearing, nor link variable fuzzing, was done for the heavy,
static, quark fields in order to preserve spatial locality,
i.e. the $\delta$ factor in (\ref{Hstatic}).
Thus, generically, replacing
$\mathcal{O} \rightarrow \mathcal{O}^{\{k\}} =
\mathcal{O}[\psi^{\{k\}},\bar{\psi}^{\{k\}}\ldots]$
the correlator (\ref{C}) becomes a $K\times K$ matrix
\begin{equation}\label{CKK}
C^{\,kk^\prime}(t,t_0)(\vec{r};t,t_0)=
\langle\mathcal{O}{}^{\{k\}}_{\mu}(\vec{r};t)\,
\bar{\mathcal{O}}{}^{\,\{k^\prime\}}_{\mu}(\vec{r};t_0)\rangle\,,
\end{equation}
with $k,k^\prime=1\ldots K$ and a sum over $\mu$ is understood.
The expression for $C^{\,kk^\prime}(t,t_0)$ in terms of quark propagators still
has the form given by (\ref{C}), however, light-quark propagator elements
are replaced with smeared ones, $G\rightarrow G^{\{\,kk^\prime\}}$.
Smearing and fuzzing prescriptions are the same at source and sink.
Thus the correlation matrix (\ref{CKK}) is hermitian by construction.

We choose an asymmetric and anisotropic lattice with geometry
$L_1\times L_2\times L_3\times L_4 = 8\times 8\times 32\times 16$.
The bare lattice constants, in the respective directions, satisfy
$a_1=a_2=2a_3=2a_4$.
Subsequently we shall refer to $a_4=a_3=a$ as the lattice constant $a$ of
the simulation, so the physical volume is $16a \times 16a \times 32a \times 16a$.
The lattice has a fine resolution in 4-direction (time), and the same fine resolution
in the (spatial) 3-direction where the lattice size
is twice as long as in the 1,2-directions.
The idea here is to place the static quark sources along the 3-direction
and thus achieve a fine spatial resolution of the relative distance as well as to
provide enough space to allow for asymptotic separation of the hadrons.

\begin{figure}[h]
\includegraphics[angle=0,width=45mm]{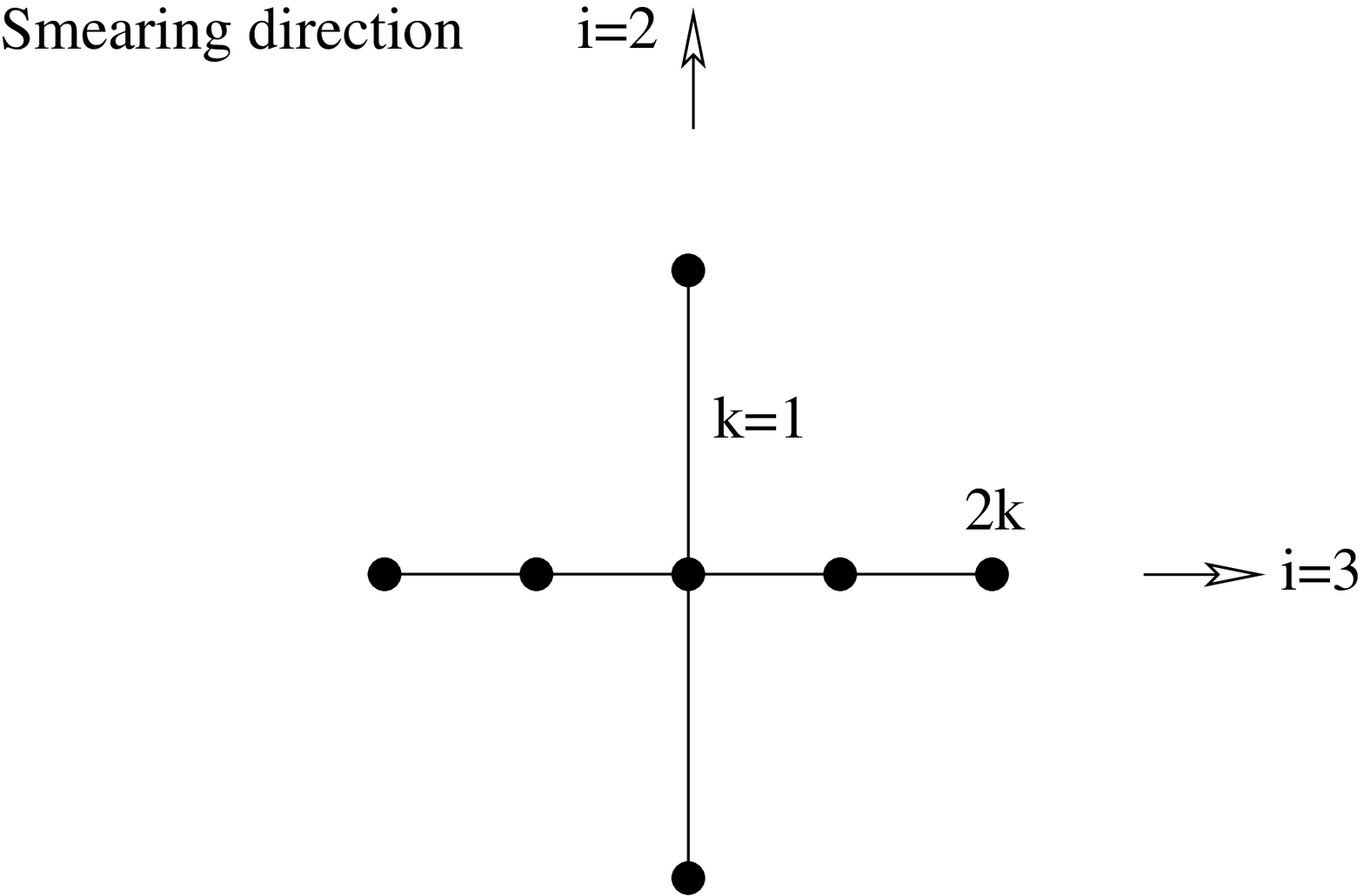}\\
\includegraphics[angle=0,width=45mm]{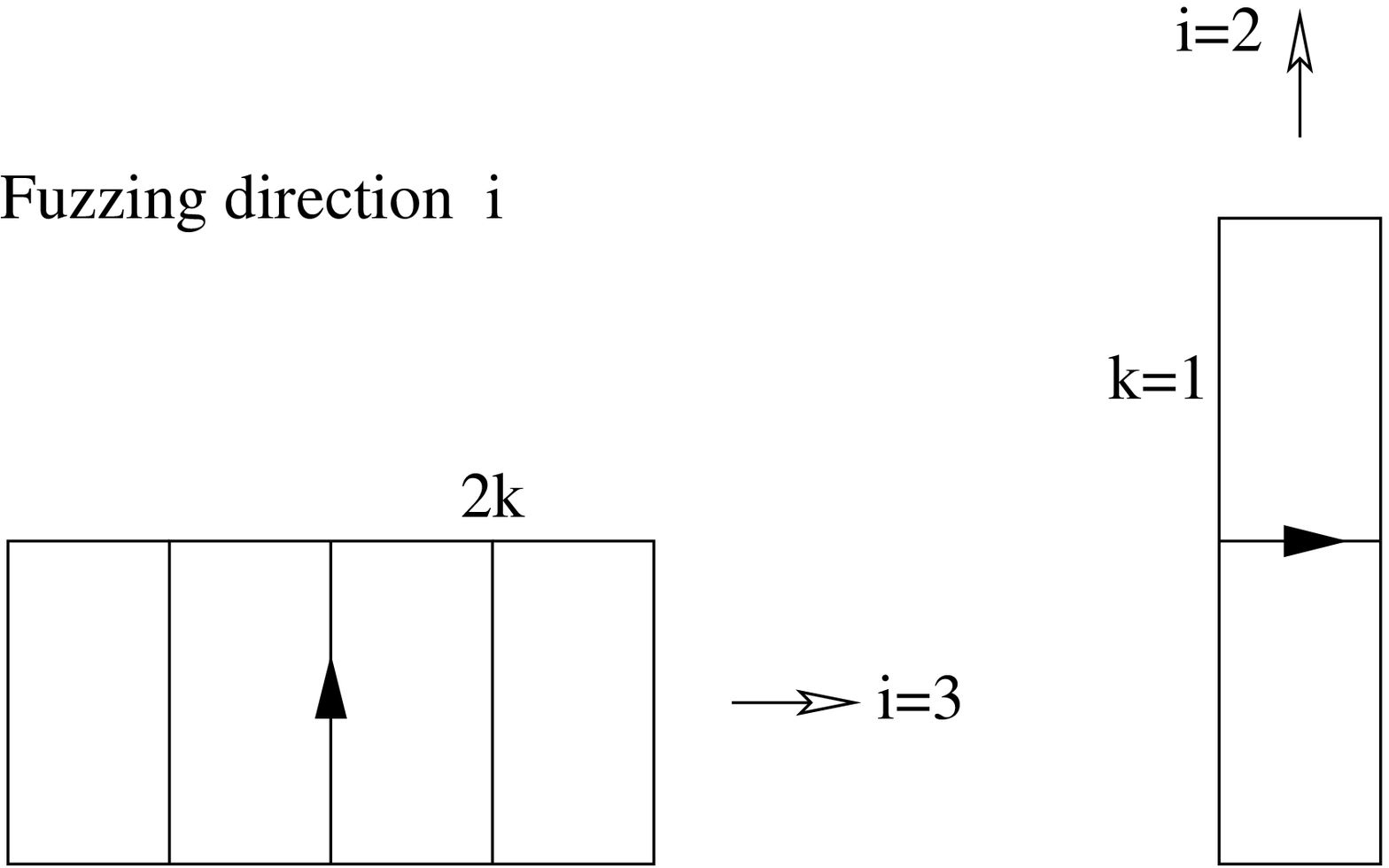}\\
\caption{Illustration of the prescription for quark field smearing (top) and gauge
field fuzzing (bottom) on the asymmetric lattice. The geometry for $k=1$
iteration is shown.}
\label{fig:asym}
\end{figure}

The spatial lattice asymmetry leads us to modify the procedure for operator
smearing. A smearing iteration of $k$ steps is understood to mean $k$ steps
in the 1,2-directions but $2k$ steps in the 3-direction, so the physical
smearing extent is the same for all spatial directions (neglecting
renormalization effects). Mutatis mutandis, the same is true for gauge field
fuzzing. An illustration is shown in Fig.~\ref{fig:asym}.
The implementation used here is described in more technical terms
in reference \cite{Cook:2006hh}.

\begin{figure}[h]
\includegraphics[angle=0,width=85mm]{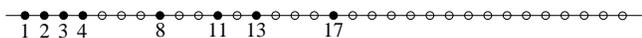}
\caption{Placement of static hadron sources along the
3-direction of the lattice. Periodic boundary conditions are applied to the
quark and gluon fields.
All distances between $r=1a$ and $r=16a$ can be achieved, some multiple times.}
\label{fig:rdistcirc}
\end{figure}

The positions of the static quark sources are chosen as
$x=(5,5,n,3)$ with $n=1,2,3,4,8,11,13,17$.
Their spatial locations along the 3-direction, see Fig.~\ref{fig:rdistcirc},
allow us to achieve any relative distance $r=1a\ldots 16a$ by choosing an
appropriate pair of source points.
Some relative distances can be obtained more than once.
In those cases appropriate averages over sets of source and sink pairs have
been taken to compute the correlator matrix elements.
Because periodic boundary conditions are applied in the spatial directions
relative distances larger than $r=16a$ are redundant.

Given the approximations employed in this work, e.g. static heavy quarks,
and a few more mentioned below, it would be of no advantage to employ
sophisticated improved lattice actions.
We have therefore used the standard Wilson plaquette action with Wilson fermions
in a quenched simulation \cite{Mon94}.
The gauge field couplings in the $\mu$-$\nu$ planes and the hopping parameters in directions
$\mu$ are given by, respectively,
\begin{equation}
\beta_{\mu\nu}=\beta\,\frac{a_1a_2a_3a_4}{(a_{\mu}a_{\nu})^2}
\quad\mbox{and}\quad
\kappa_\mu = \frac{\kappa}{a_\mu\frac{1}{4}\sum_{\nu=1}^{4}\frac{1}{a_\nu}}\,.
\end{equation}
The simulation was done at $\beta=6.2$ with four values 
$\kappa=0.140,\, 0.136,\, 0.132,\, 0.128$ of the hopping parameter employing a
multiple mass solver \cite{Glassner:1996gz}.
A total of 343 gauge field configurations were used.

\section{\label{sec:ana}Mass spectrum analysis}

The time evolution of the eigenvalues of $C(t,t_0)$ determines the energy spectrum.
Eigenvalues, behaving exponentially with $t$,
may rapidly vanish into numerical noise. Conventional diagonalization methods do not work
well under those circumstances. Singular value decomposition (SVD), on the other hand,
is ideally suited to the task \cite{Golub96}. The SVD reads
\begin{equation}\label{svd}
C(t,t_0)=U(t,t_0)\,\Sigma(t,t_0)\,V^{\dagger}(t,t_0)\,,
\end{equation}
where $U(t,t_0)$ and $V(t,t_0)$ are unitary in our case\footnote{This decomposition
easily generalizes to the case of a different number of operators at source and sink.},
and $\Sigma(t,t_0)={\rm diag}(\sigma_{1}(t,t_0)\ldots\sigma_{K}(t,t_0))$ contains the
singular values satisfying $\sigma_{k}(t,t_0)\ge 0$.
If $C(t,t_0)$ is non-degenerate and positive semi-definite then the set
of singular values $\{\sigma_k(t,t_0):k=1\ldots K\}$ and the set of eigenvalues
are the same. A few more details can be found in \cite{Cook:2006hh}.
For simplicity we will refer to $\sigma_k(t,t_0)$ as eigenvalues.

To extract energy levels from the eigenvalues an often used procedure,
sometimes starting from a generalized
eigenvalue problem \cite{Luscher:1990ck}, is to construct effective mass functions
from the eigenvalues and then try to identify plateaus in the asymptotic time
region. Typically, only a narrow subset of the time slices is usable, which also is
subject to some discretion.
We shall not rely on effective mass functions here, but rather analyze
correlator eigenvalues by means of the maximum entropy method
\cite{Jar96} following the adaption described in \cite{Cook:2006hh},
and in \cite{Fiebig:2002sp}.
Among the advantages are that all available time slices may be used, if desired,
and that excited states, if present, are easily revealed.

The time dependence of the eigenvalues $\sigma_{k}(t,t_0)$ shall be fit with
the model 
\begin{equation}\label{Fmodel}
F(\rho|t)=\int_{0}^{\infty}d\omega\,
\rho(\omega)e^{-\omega t}\,.
\end{equation}
for some set of discrete times slices $t=t_1\ldots t_2$, where $\rho(\omega)$ is a
spectral density function.
The maximum entropy method (MEM) is based upon Bayesian statistics.
In this context the numbers $\rho(\omega)$ are interpreted as stochastic
variables \cite{Jar96}. Their most likely values are obtained by simulated annealing as
described in \cite{Cook:2006hh}. We refer the reader to this reference for
technical details.

\begin{figure}[h]
\includegraphics[angle=0,width=64mm]{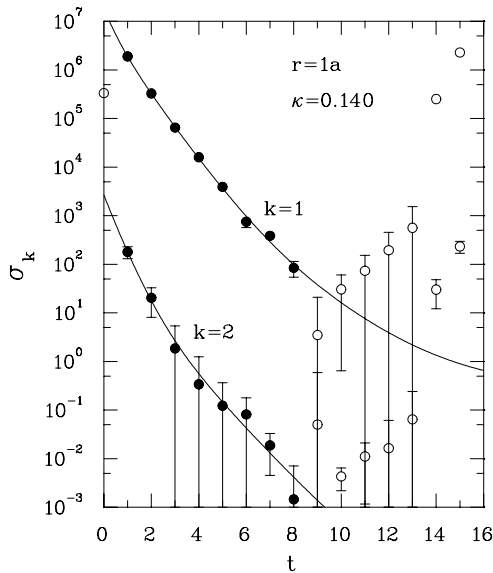}\noindent\rule{10mm}{0mm}
\caption{Two sets of eigenvalues $\sigma_k(t,t_0), k=1,2,$ of the $3\times 3$
time correlation matrix (\protect\ref{svd}) for relative distance $r=1a$ and
quark hopping parameter $\kappa=0.140$. The third eigenvalue, $k=3$, is not shown,
it vanishes into numerical noise. Filled plot symbols belong to time slices
used for the MEM fits employing the model (\protect\ref{Fmodel}),
the curves are the corresponding results.
Note that the source has been shifted to $t_0=0$.}
\label{fig:corval}\end{figure}

At each of the $16$ relative distances $r$ we have used three ($K=3$) operator smearing
levels $k=1,2,3$. A typical set of eigenvalues is displayed in Fig.~\ref{fig:corval}
for distance $r=1a$ and the smallest (light) quark mass, $\kappa=0.140$.
Only two eigenvalue correlators are shown, the third one is indistinguishable from
zero due to numerical noise. Note that eigenvalues are separated by about four
orders of magnitude. Thus crossing ambiguities are nonexistent\footnote{We have
found that solving the generalized eigenvalue problem, as put forward
in \protect\cite{Luscher:1990ck}, does not promote this desirable feature.}.
The eigenvalue sets for all other distances $r$ exhibit the same features.

\begin{figure}
\includegraphics[angle=90,width=68mm]{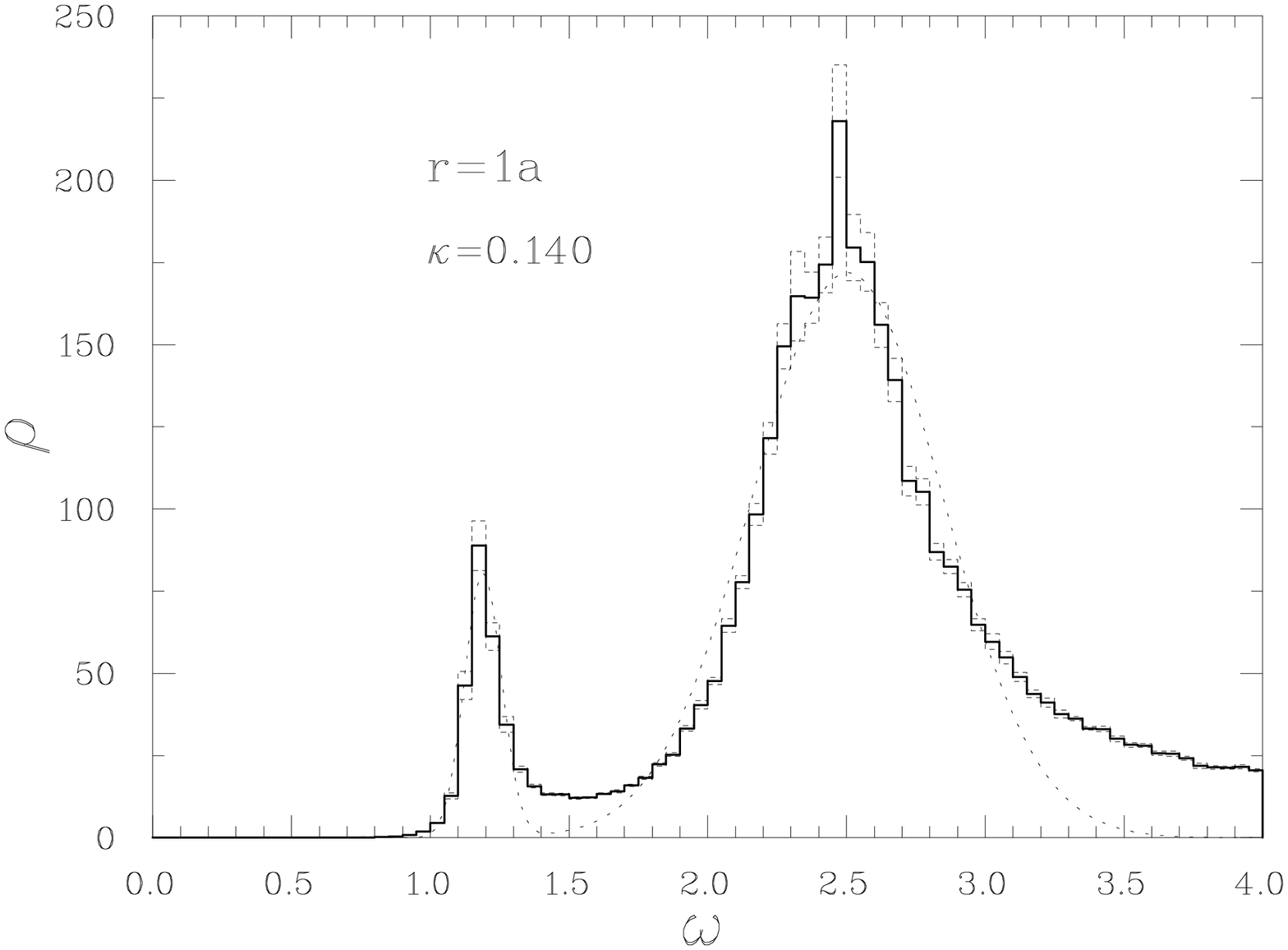}\hspace{1mm}\vspace{2mm}\\
\includegraphics[angle=90,width=70mm]{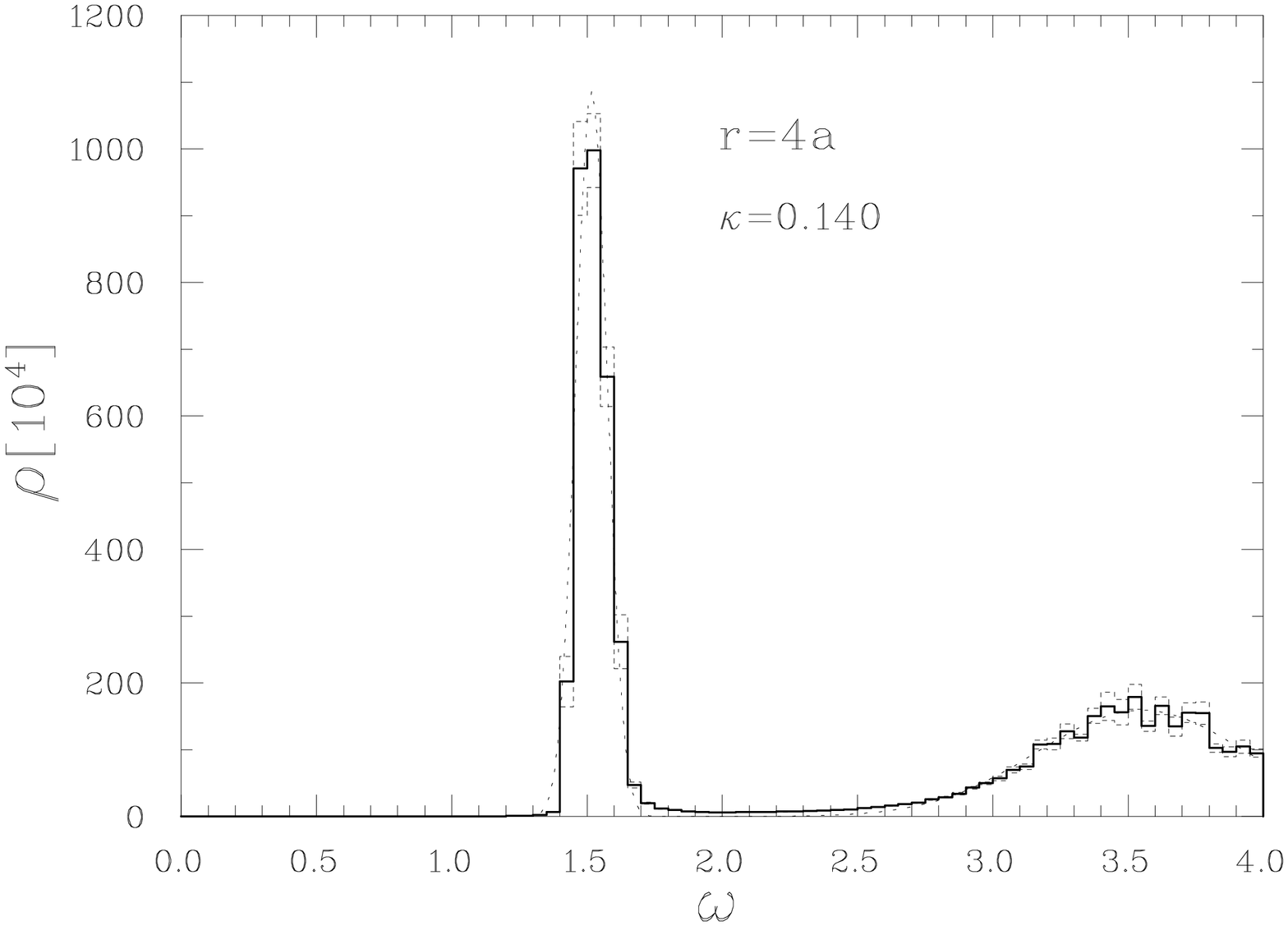}\hspace{2mm}\vspace{2mm}\\
\includegraphics[angle=90,width=68mm]{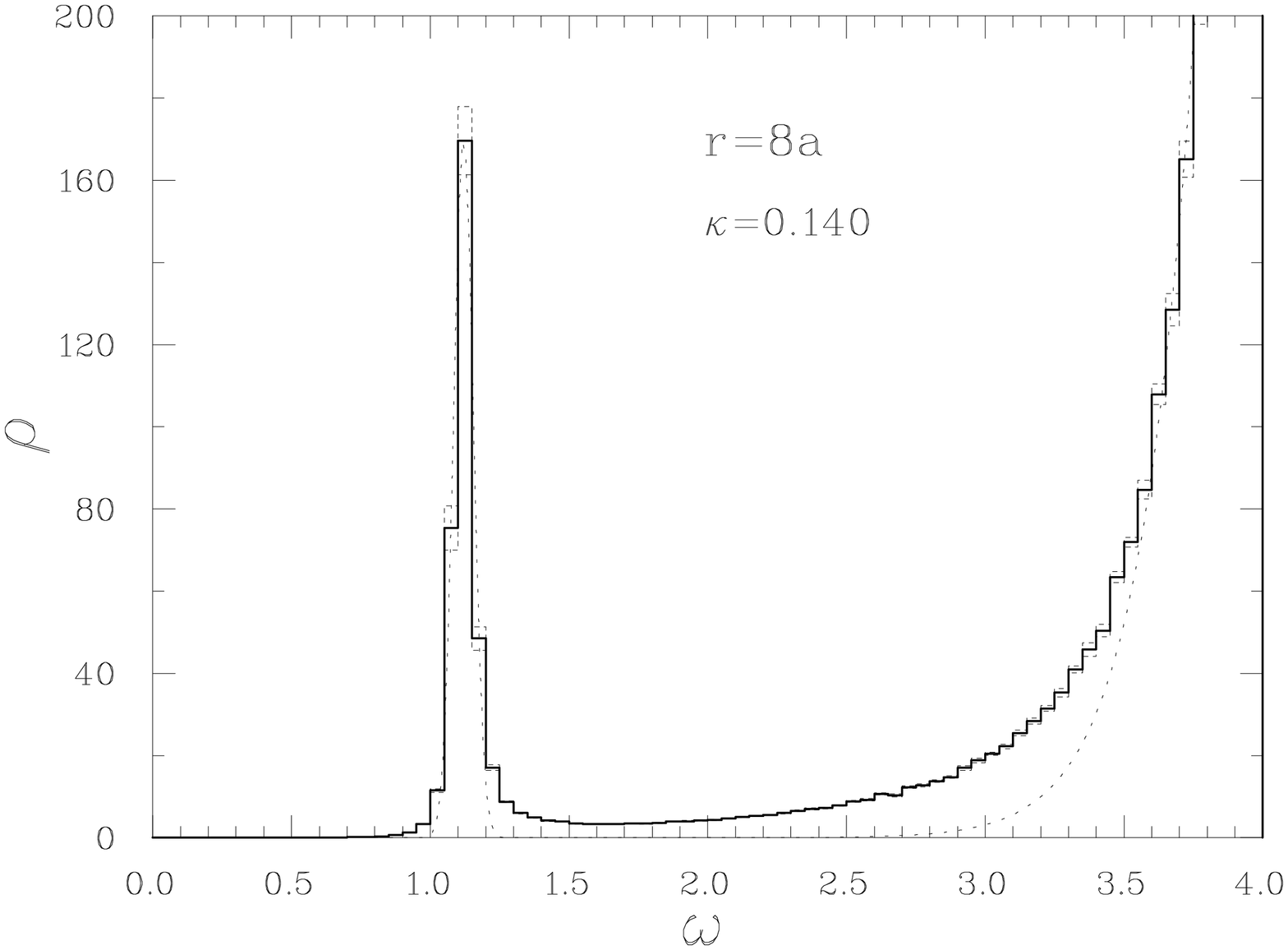}\hspace{0mm}\vspace{2mm}\\
\includegraphics[angle=90,width=69mm]{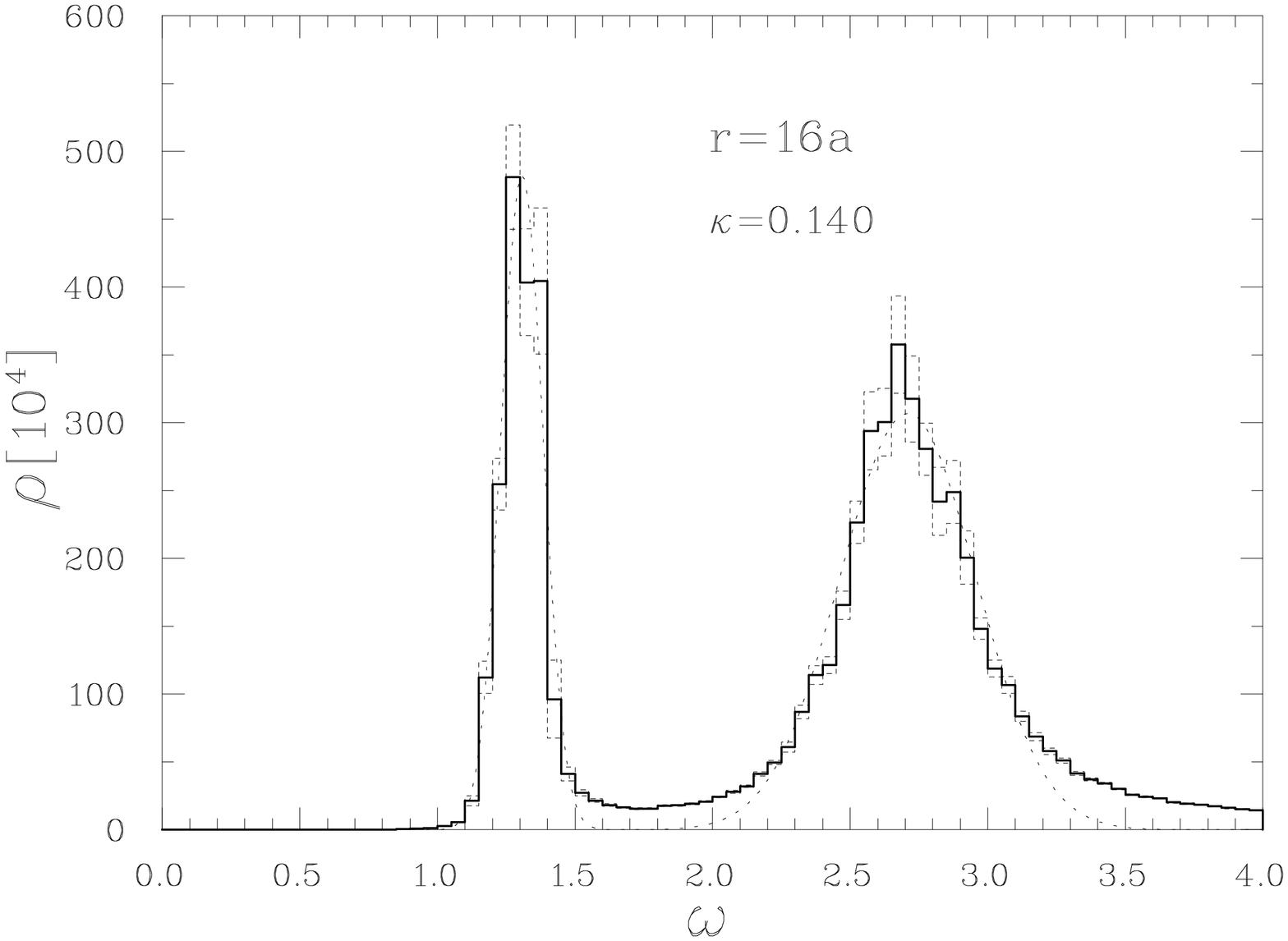}\hspace{0mm}
\caption{Spectral density functions $\rho$, according to (\protect\ref{Fmodel}),
for two-hadron operators with relative distance $r=1,4,8,16$, in units of $a$,
thick histogram lines. Shown are eigenvalue spectra which contain ground state peaks.
The dashed histogram lines indicate errors as explained in the text.
The smooth dotted lines correspond to Gaussian fits to the discretized spectral densities.}
\label{fig:memspec}\end{figure}

The eigenvalue correlators were fit with the model (\ref{Fmodel}) using,
consistently in all cases, time slices $t=1\ldots 8$ with the source $t_0=0$
being omitted, and mass range of $0\leq\omega\leq 4$ with
discretization interval size $\Delta\omega=0.05$, all in units of $a^{-1}$.
The maximum entropy method then yields spectral density functions $\rho(\omega)$.
Representative examples are shown in Fig.~\ref{fig:memspec} for $r=1,4,8,16$,
in units of $a$, and $\kappa=0.140$.
The histogram lines in Fig.~\ref{fig:memspec} represent $\rho(\omega)$,
obtained from simulating annealing (cooling) following the very procedure
put forward in \cite{Cook:2006hh}. Sixteen random annealing starts were used. 
The solid histogram lines represent the average of those 16 runs, the dashed
histogram lines give the corresponding standard deviation.
In most cases the spectra exhibit isolated peaks, say 
$\delta_n=\{\omega:\omega\in{\rm peak}\ \#n\}$.
Then one may compute, for each peak $n$, the volume $Z$,
the energy $E$, and the width $\Delta$, according to
\begin{eqnarray}
Z&=&\int_{\delta_n}d\omega\/\rho(\omega)\label{Zn}\\
E&=&Z^{-1}\int_{\delta_n}d\omega\/\rho(\omega)\omega\label{En}\\
\Delta^2&=&Z^{-1}\int_{\delta_n}d\omega\/\rho(\omega)\left(\omega-E\right)^2\,.
\label{Dn}\end{eqnarray}
Typically the largest eigenvalues reveal the ground state. There are few
exceptions. The spectra at $r=1a$ and $r=8a$, shown in Fig.~\ref{fig:memspec},
are examples.
Selecting the support $\delta_n$ of a peak can be slightly ambiguous, particularly
if the peak sits on background noise, or if there is overlap with another one.
We have therefore, as a matter of course, performed least-square fits to the
average spectral functions $\rho(\omega)$ with one or two Gaussians, as required
in each case. These fits then give values for $Z$, $E$,
and $\Delta$ as defined in (\ref{Zn})--(\ref{Dn}) directly from the
parameters of the Gaussians. 
In this manner we have obtained 16 sets, one for each $r$, of five-quark hadron
ground state energies. Each set contains four energies corresponding to hopping
parameter values $\kappa=0.140,\, 0.136,\, 0.132,\, 0.128$.

The spectral density functions $\rho(\omega)$ typically exhibit some
additional structure at the high end of the $\omega$ range, a broad
peak in most cases. These secondary peaks appear because diagonalizing
the correlator matrix separately on all time slices only ensures that
its eigenvalues describe a single state from the physical spectrum at
large values for $t$ \cite{Luscher:1990ck}. At small time slices,
close to the source, the eigenvectors of $C(t,t_0)$ may still describe
a mix of spectral states. In practice the large-$t$ rule means
that, in each eigen channel, one should only take
the lowest peak into account. In physical terms the secondary peaks are
separated from those by $\approx$3GeV, and larger, and thus should
probably be considered lattice artifacts.
 
\begin{figure}
\noindent\rule{1mm}{0mm}
\includegraphics[angle=90,width=76mm]{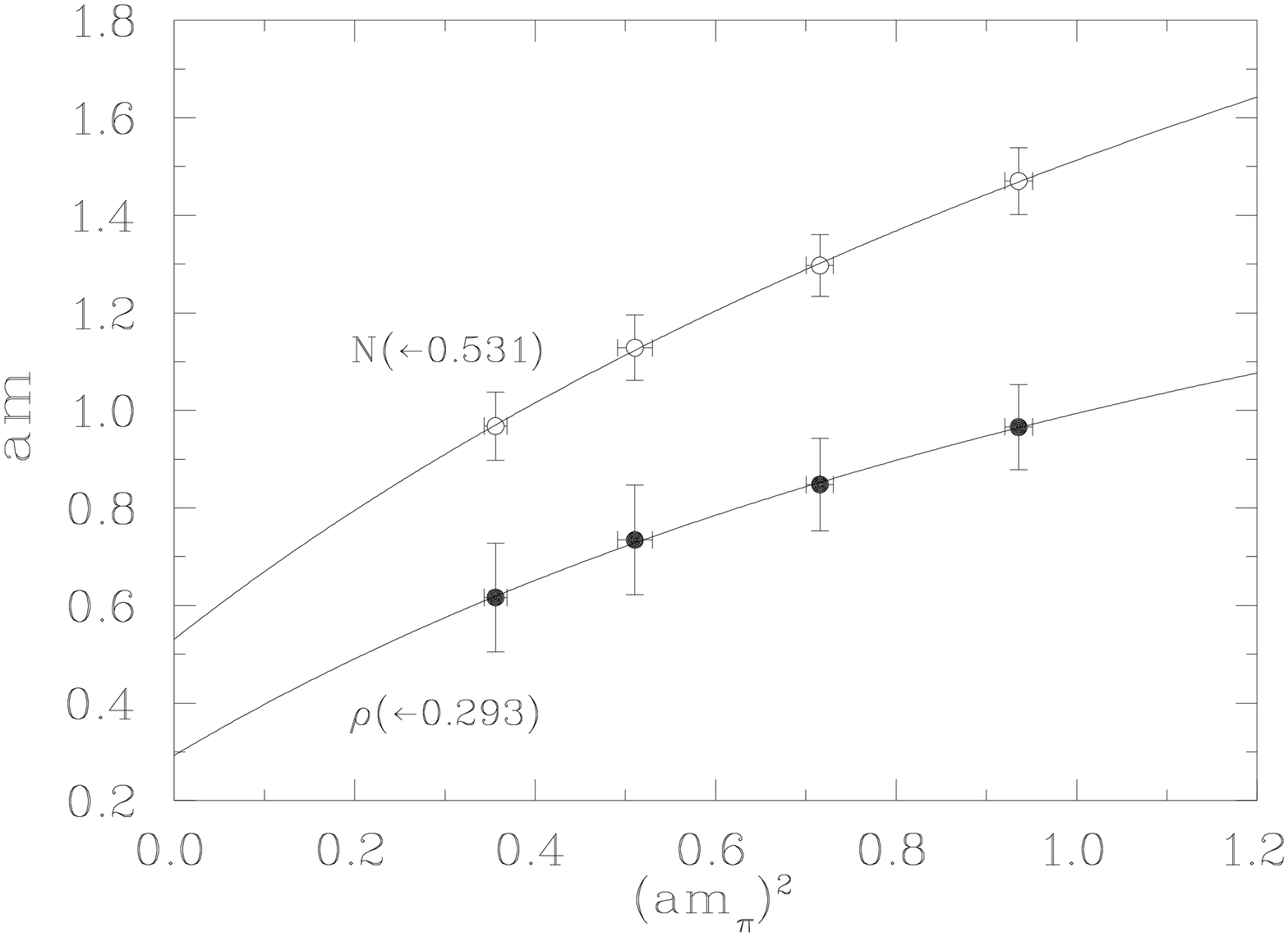}\hspace{4mm}\\
\includegraphics[angle=90,width=77mm]{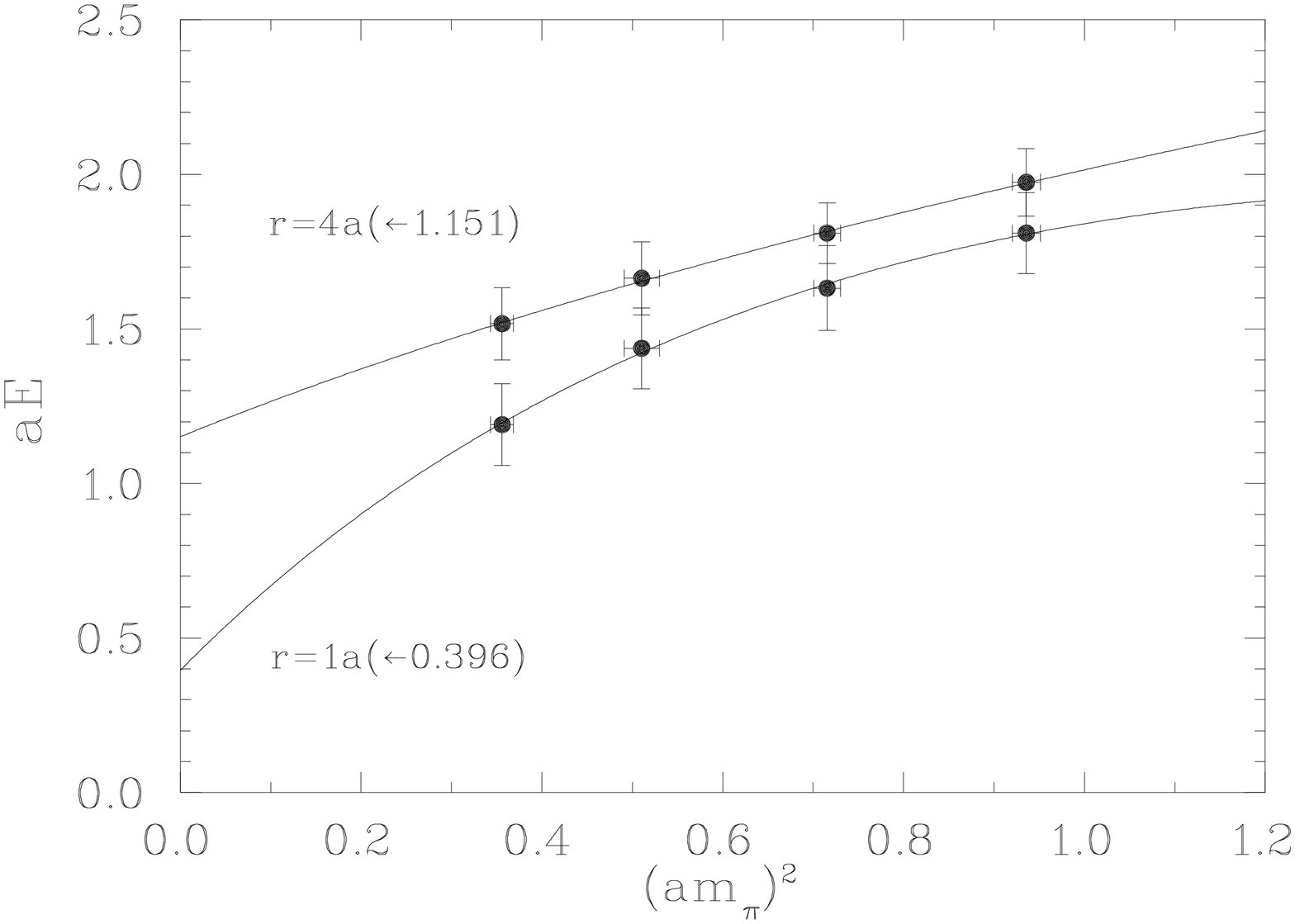}\hspace{0mm}
\caption{Extrapolations of the vector meson and nucleon masses $y=am$
based on the model (\protect\ref{mpi0}) and four hopping
parameter values, top panel.
The bottom panel shows two typical examples of five-quark hadron ground state
energies $y=aE$, at distances $r=1a$ and $r=4a$, and their respective
extrapolation using the same procedure.}
\label{fig:extra}\end{figure}

In order to relate our results to the physical hadron spectrum we have also
computed $3\times 3$ correlation matrices with standard local \cite{Mon94}
pseudoscalar meson ($\pi$),
vector meson ($\rho$), and nucleon ($N$) operators,
using the same smearing and fuzzing prescription. The analysis was
performed with the MEM in exactly the same way as described above.
This now allows extrapolations of hadron masses to the physical pion
mass region ($am_\pi\approx 0$).
We use the extrapolation model introduced in \cite{Cook:2006hh}
\begin{equation}
y=c_1+c_2x+c_3\ln(1+x)\,,\quad x=(am_\pi)^2\,,
\label{mpi0}\end{equation}
and $y=am$.
For a motivation of the same we refer the reader to said reference.
The extrapolation prescription is, of course, a source of systematic
error on our results.
In Fig.~\ref{fig:extra} we show the vector meson and nucleon masses
as a function of the squared pion mass and the extrapolations according
to (\ref{mpi0}).
The extrapolated values $am_\rho=0.293$ and $am_N=0.531$, as
$x\rightarrow 0$, are used to set the reduced mass of the $\rho N$ system to its
experimental value $m_{\rho N}=424.7$MeV.
This results in a lattice constant of $a^{-1}=2251$MeV or $a=0.088$fm.
The corresponding vector meson and nucleon masses come out as
659MeV and 1196MeV, respectively, which deviate by $15\%$ and $27\%$ from
their experimental values. We take these numbers as indicators of the
systematic errors to be expected due to the extrapolation.

\begin{figure}
\hspace{-1mm}
\includegraphics[angle=90,width=85mm]{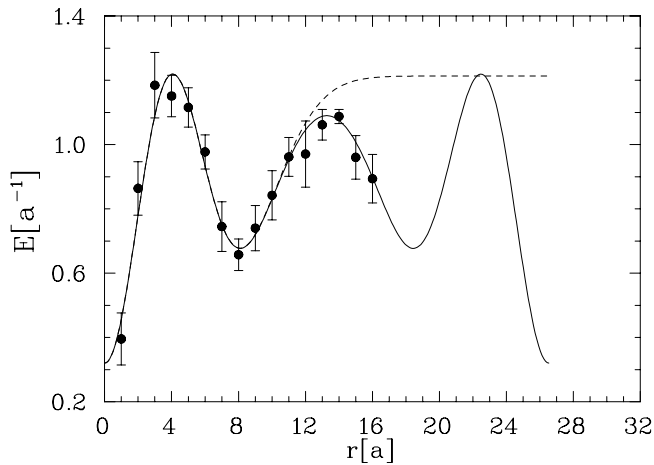}\hspace{2mm}
\caption{Extrapolated ground-state energy $E$ of the heavy-light $K$--$\Lambda$
like system versus the relative distance $r$ with MEM based uncertainties
(plot symbols). The solid line
represents a fit with the model (\protect\ref{VL}). The dashed line shows
the same fit, but the mirror term $aV(L-x)$ has been dropped.}
\label{fig:Wpot}\end{figure}

Also in Fig.~\ref{fig:extra} two representative examples of
extrapolations of five-quark hadron ground state masses $y=aE$ are shown,
the relative distances are $r=1a$ and $r=4a$. The error bars
represent spectral peak widths $\Delta$ according to (\ref{Dn}).
We utilize those to compute uncertainties on the extrapolated energies:
At each of the four data points stochastically independent Gaussian
random numbers with average and standard deviation given by the energies
and peak widths of the data, respectively,
were generated and again fitted with (\ref{mpi0}). Repeating this a large
number of times then gives rise to a
standard deviation which we interpret as a MEM peak extrapolated width $\Delta$.
The extrapolated energy spectrum obtained in this way is displayed
in Fig.~\ref{fig:Wpot} as a function of the relative distance $r$.

The energies shown in Fig.~\ref{fig:Wpot} are also subject to statistical
errors which originate from the gauge configuration ensemble.
To identify those a bootstrap error analysis \cite{Efron:bootstrap} was performed.
Using a resampling factor of eight on the 343 gauge configurations the MEM
spectral analysis was repeated for each bootstrap sample\footnote{In principle
jackknife samples can also be used but the amount of effort would be prohibitive.}.
In each case the same
fixed random number sequence was employed in order to eliminate fluctuations
in the annealing process. The statistical errors $S$ thus obtained are
compared to the peak widths $\Delta$ in Tab.~\ref{tab:Serr}.
Results are for $\kappa=0.140$ (lightest quark mass).
The other $\kappa$ values give very similar numbers because the underlying
set of gauge configurations is the same.
Clearly the statistical errors $S$ are consistently smaller than the peak
widths $\Delta$. On average over all relative distances $r$ the ratio
$\Delta/S$ turns out to be slightly larger than two.
Thus, judiciously keeping the larger errors, we have used the spectral peak widths
as the principle input for computing uncertainties.
\begin{table}[h]
\caption{\label{tab:Serr} Comparison of the statistical errors $S$ of
the ground-state peak positions $E$, see (\protect\ref{En}), and the
peak widths $\Delta$, see (\protect\ref{Dn}).
Results are shown for the lightest quark mass, $\kappa=0.140$, and for
all relative distances $r$.}
\begin{ruledtabular}
\begin{tabular} {rccc}
$r/a$ & $aE$ & $a\Delta$ & $aS$ \\ \hline
   1 & 1.193 & 0.050 & 0.038 \\
   2 & 1.434 & 0.106 & 0.031 \\
   3 & 1.514 & 0.083 & 0.037 \\
   4 & 1.526 & 0.062 & 0.059 \\
   5 & 1.512 & 0.067 & 0.076 \\
   6 & 1.378 & 0.053 & 0.032 \\
   7 & 1.206 & 0.052 & 0.024 \\
   8 & 1.120 & 0.035 & 0.029 \\
   9 & 1.190 & 0.055 & 0.030 \\
  10 & 1.223 & 0.065 & 0.020 \\
  11 & 1.348 & 0.066 & 0.034 \\
  12 & 1.390 & 0.072 & 0.030 \\
  13 & 1.465 & 0.047 & 0.024 \\
  14 & 1.338 & 0.080 & 0.030 \\
  15 & 1.315 & 0.058 & 0.040 \\
  16 & 1.297 & 0.062 & 0.009 
\end{tabular}
\end{ruledtabular}
\end{table}

\section{Adiabatic potential}

\begin{figure}[h]
\hspace{-1mm}
\includegraphics[angle=90,width=85mm]{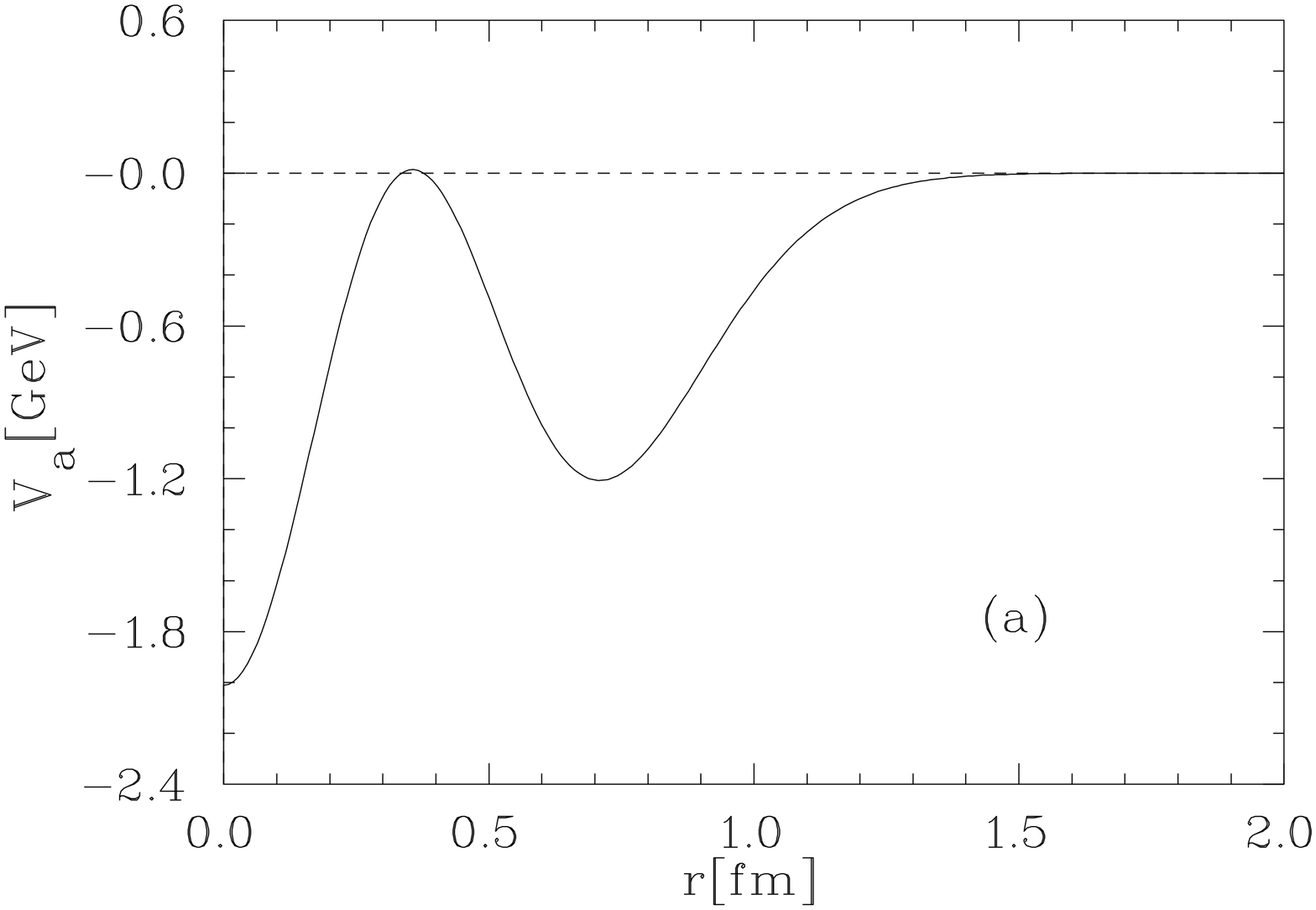}\hspace{2mm}\vspace{3mm}\\
\includegraphics[angle=90,width=85mm]{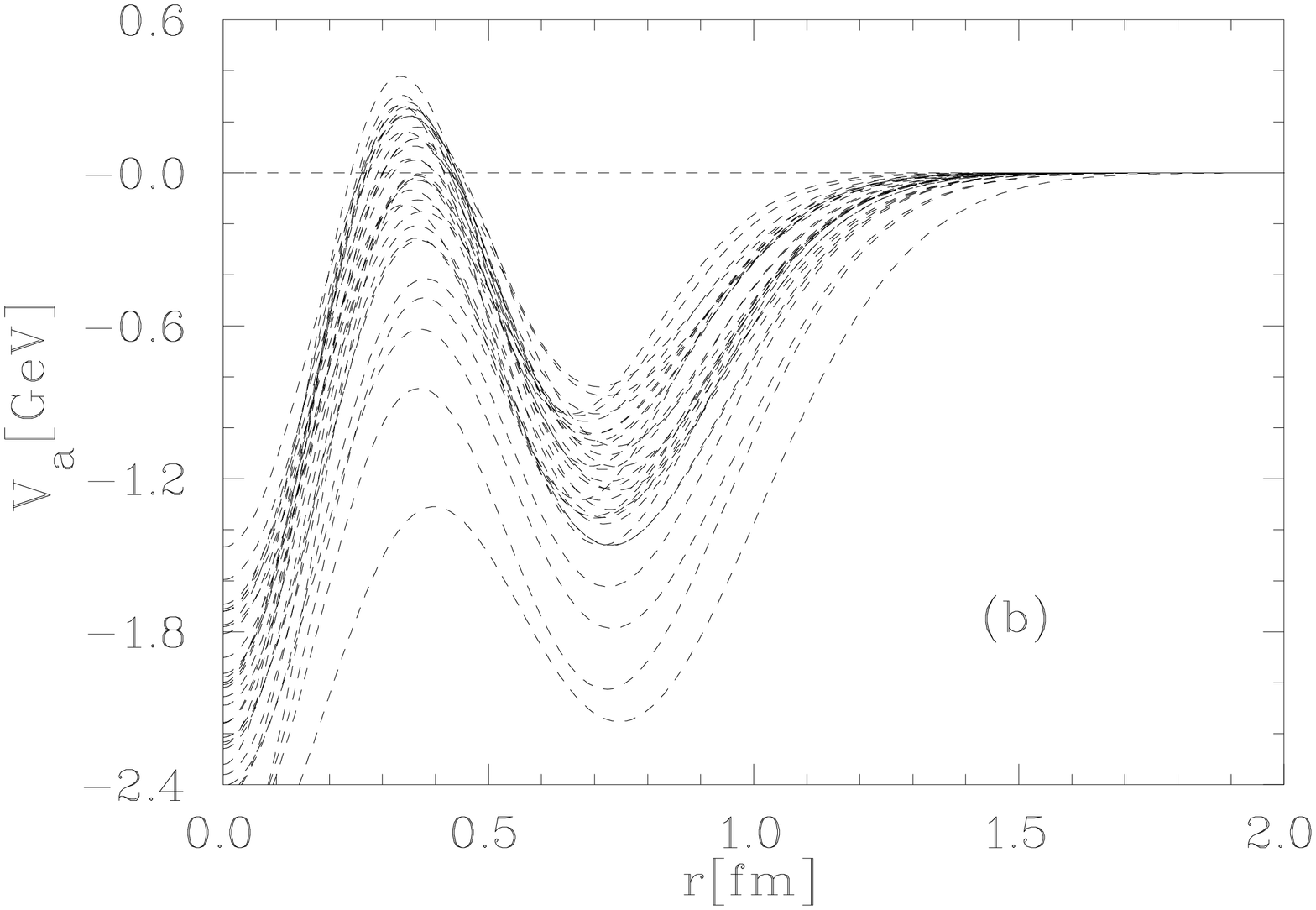}
\caption{Adiabatic potential $V_a(r)$ resulting from the lattice simulation
in physical units, panel (a). Panel (b) shows potentials from 32 sets of randomizations
of the ground state energies $E$ as seen in Fig.~\protect\ref{fig:Wpot}.}
\label{fig:Vpot}\end{figure}

The data points in Fig.~\ref{fig:Wpot} connect smoothly along a
continuous curve. In order to find an appropriate fit model we note that
the ground-state $K$--$\Lambda$ like system should be in a relative S-wave.
Our choice thus are S-wave 3d harmonic oscillator functions \cite{Talman:1970}.
The first few terms are given by
\begin{equation}\label{V}
aV(x)=\exp(-\alpha_1x^2)(\alpha_2+\alpha_3x^2+\alpha_4x^4)\,,\quad x=r/a\,,
\end{equation}
with parameters $\alpha_1\ldots\alpha_4$.
This will not be sufficient though for two reasons.
First, periodic boundary conditions have been imposed on the lattice in the 3-direction.
Thus lattice sites with relative distances $x$ and $L_3-x$ are equivalent. However, as
the two-hadron operators for the $K$--$\Lambda$ like system are essentially loop operators,
see Fig.~\ref{fig:CHG}, those do not share the periodic behavior of the fundamental
lattice fields.
This effect should be most notable for distances around $\approx L_3/2=16$.
Second, interactions with hadron images in adjacent copies of the lattice are present.
Their dominant effect is a constant interaction energy, as the relative distances of
most of the mirror images remain constant with varying $x$.
Thus motivated, empirically one finds that the lattice data are well represented by
the model
\begin{equation}\label{VL}
aV_L(x)=\alpha_0+aV(x)+aV(L-x)\,,\quad x=r/a\,,
\end{equation}
where $\alpha_0$ and $L$ are additional parameters.
The resulting fit is shown in Fig.~\ref{fig:Wpot} as a solid line.
The dashed line represents $\alpha_0+aV(x)$ only.
Table~\ref{tab:Wparam} contains the fit parameters. 
\begin{table}[h]
\caption{\label{tab:Wparam} Results for the fit parameters using the model (\protect\ref{VL}).}
\begin{ruledtabular}
\begin{tabular}{cd}
$\alpha_0$ & 1.213(93)    \\
$\alpha_1$ & 0.0409(37)   \\
$\alpha_2$ &-0.89(10)     \\
$\alpha_3$ & 0.109(19)    \\
$\alpha_4$ &-0.00328(52)  \\
$L$        &26.54(57)
\end{tabular}
\end{ruledtabular}
\end{table}
Note that $L$ deviates somewhat from $32$, as should be expected for the reason
stated above.
The $\chi^2$ per degree of freedom is $0.63$, the uncertainties given
stem from the covariance matrix of the Levenberg-Marquardt algorithm.
We list those for completeness only, they will not be used in the subsequent analysis.
Rather, because we are mostly interested in uncertainties of possible bound
state energies, as the case may be, we proceed as follows:
Each data point in Fig.~\ref{fig:Wpot} is randomized 
by stochastically independent Gaussian
random numbers with average and standard deviation given by the energies
and peak widths of the data, respectively. Then the fit with the model (\ref{VL}) is
repeated, the corresponding set of parameters gives rise to a new potential
$V(x)$. The procedure is performed 32 times.
We show the results in Fig.~\ref{fig:Vpot} along with the potential obtained
directly from the original data points. Specifically, Fig.~\ref{fig:Vpot}
displays adiabatic potentials
\begin{equation}\label{Va}
V_a(r)=V(r/a)
\end{equation}
on physical scales given by the lattice
constant $a$, as determined above.

\section{Bound states and wave functions}

Figure~\ref{fig:Vpot} represents the end result of the lattice simulation.
Although we expect that systematic errors coming from extrapolations may
be as large as $30\%$, and the uncertainties on the potential itself according
to Fig.~\ref{fig:Vpot}(b) are sizable, we believe that important physics can
still be learned on a qualitative level.
First, we observe that $V_a(r)$ of Fig.~\ref{fig:Vpot}(a) has no repulsive regions.
This is a reasonable, and confidence inspiring,
outcome because repulsion in a two-hadron system is typically Pauli repulsion,
but of course the latter is absent in a $K$--$\Lambda$ like system.

Also, as mentioned earlier just before Eqn.~(\ref{C}), the relative distance $r=0$ was excluded
from the simulation because ``color twisted'' operators would then have to be
included for all distances $r$, to be consistent. 
Without further work it remains an open question how such a family of operators might
influence the form of $V_a(r)$. However, it seems reasonable to assume that those
will not play a major role for large $r$ because there they would create hadrons
inconsistent with confinement, whereas at $r=0$ mixing with those states would lower
the ground state energy according to general principles of quantum mechanics.
Thus, we believe that the current simulation correctly captures the most important
physics of the four-quark system.

\begin{figure}[h]
\center
\includegraphics[angle=90,width=82mm]{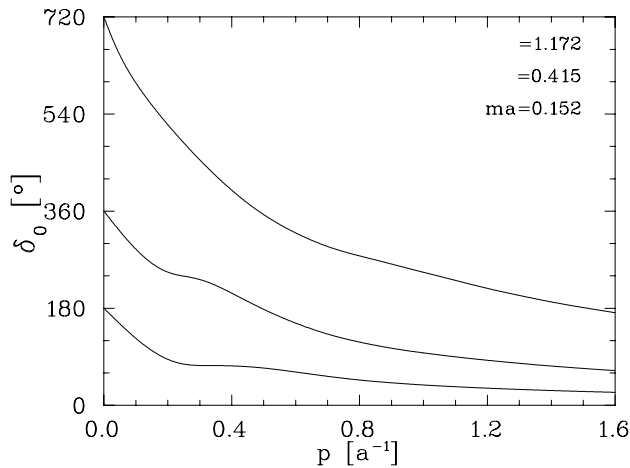}
\caption{Scattering phase shifts (S-wave) for the $K$--$\Lambda$, $D$--$\Lambda_c$,
and $B$--$\Lambda_b$ systems (bottom to top, respectively) versus the
relative momentum $p$.}
\label{fig:scatt}\end{figure}

The salient features of $V_a(r)$ are the two distinct regions of
attraction around $r\sim 0.5$--$1.0$fm and $r\lesssim 0.2$fm.
An obvious question is whether the attraction is sufficient to generate bound
states of the five-quark molecule. To find out we have, first, computed
scattering phase shifts $\delta_\ell(p)$ employing the adiabatic
potential $V_a(r)$ in a Schr{\"o}dinger equation.
For the reduced mass $m$ we chose, in turn, the three values of the physical
$K$--$\Lambda$, $D$--$\Lambda_c$, and $B$--$\Lambda_b$ systems \cite{Yao:2006px}.
These evaluate to $am=0.152, 0.415$, and $1.172$, respectively.
Continuum boundary conditions were implemented by solving a Volterra integral
equation within the Jost function formalism \cite{Tay72}.
The resulting scattering phase shifts are shown in Fig.~\ref{fig:scatt}.
According to Levinson's theorem, see \cite{Tay72} for example,
we have $\delta_\ell(0)-\delta_\ell(\infty)=n\pi$ where $n$ is the
number of bound states. In the S-wave $n$ is one for the $K$--$\Lambda$ system,
two for the $D$--$\Lambda_c$ system, and four for the $B$--$\Lambda_b$ system.
We need to caution though that relativistic effect are large in $K$--$\Lambda$,
with increasing quark mass ($s\rightarrow c\rightarrow b$) this becomes less
of a concern. The scattering phase shifts are quite featureless, but they
give us an exact count of the number of bound states present in the simulation. 

In order to compute bound state energies and wave functions we solve the radial
Schr{\"o}dinger equation with $V_a(r)$. Our conventions can be gleaned from
\begin{equation}\label{SchrEqn}
-\frac{d^2u_a}{dr^2}+[2mV_a(r)+\frac{\ell(\ell+1)}{r^2}]u_a=2mEu_a\,,
\end{equation}
and
\begin{equation}\label{unorm}
\int_0^\infty dr |u_a(r)|^2=1\,.
\end{equation}
The region between $r=0$ and $r=30a$ was subdivided into 60 discretization intervals.
A suitable cubic B-spline representation \cite{DeBoor78} for $u_a(r)$
was employed to enforce bound-state boundary conditions and provide smooth
interpolation between the knots. This then leads to a generalized matrix
eigenvalue problem.
The scale neutral bound state energies are compiled in Tab.~\ref{tab:bstate} for
the three reduced masses of $K$--$\Lambda$, $D$--$\Lambda_c$, $B$--$\Lambda_b$,
and three partial waves. The errors come from repeating the calculation with 32
randomizations of the potential, as seen in Fig.~\ref{fig:Vpot}(b).
\begin{table}[h]
\caption{\label{tab:bstate}Scale neutral bound state energies $aE_n$ of the
adiabatic potential $V_a(r)$ for three values $am$ of the reduced mass $m$
and partial waves $\ell$. The reduced masses correspond
to $K$--$\Lambda$, $D$--$\Lambda_c$, and $B$--$\Lambda_b$.}
\begin{ruledtabular}
\begin{tabular} {cccccc}
$\ell$ & $am$ & $aE_1$ & $aE_2$ & $aE_3$ & $aE_4$ \\ \hline
0 & 0.152 & -0.22(12) &           &           &           \\
  & 0.415 & -0.32(13) & -0.04(13) &           &           \\
  & 1.172 & -0.39(13) & -0.26(16) & -0.15(12) & -0.001(88)\\ \hline
1 & 0.152 & -0.11(11) &           &           &           \\
  & 0.415 & -0.28(13) &           &           &           \\
  & 1.172 & -0.38(13) & -0.14(12) &           &           \\ \hline
2 & 0.152 &           &           &           &           \\
  & 0.415 & -0.20(13) &           &           &           \\
  & 1.172 & -0.35(13) & -0.11(12) &           &
\end{tabular}
\end{ruledtabular}
\end{table}

\begin{figure}
\center
\includegraphics[angle=90,width=72mm]{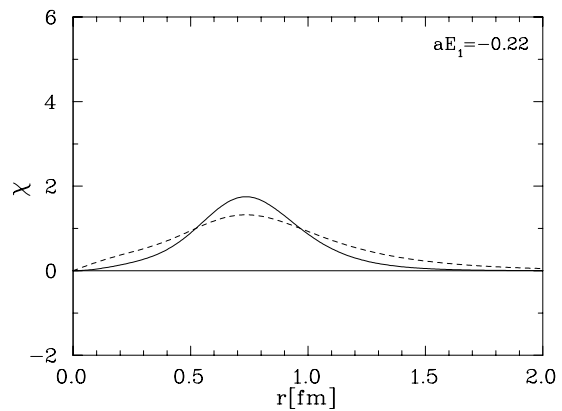}\hspace{4mm}
\caption{S-wave bound state wave function $\chi=u_a(r)$ (dashed line) and its square
$\chi=|u_a(r)|^2$ (solid line) for the $K$--$\Lambda$ system.}
\label{fig:5s}\end{figure}

\begin{figure}
\center
\includegraphics[angle=90,width=72mm]{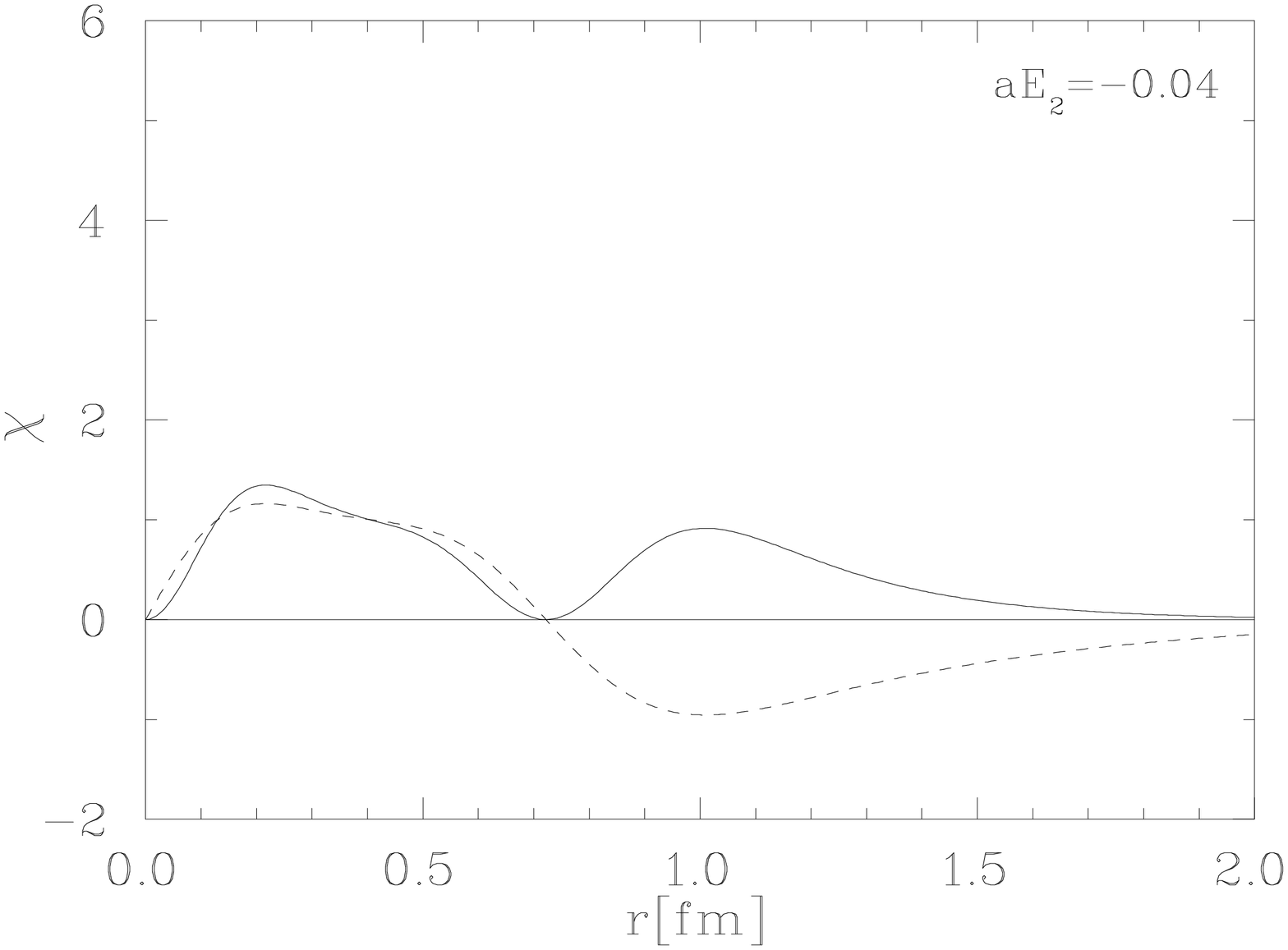}\hspace{4mm}\vspace{2.8mm}\\
\includegraphics[angle=90,width=72mm]{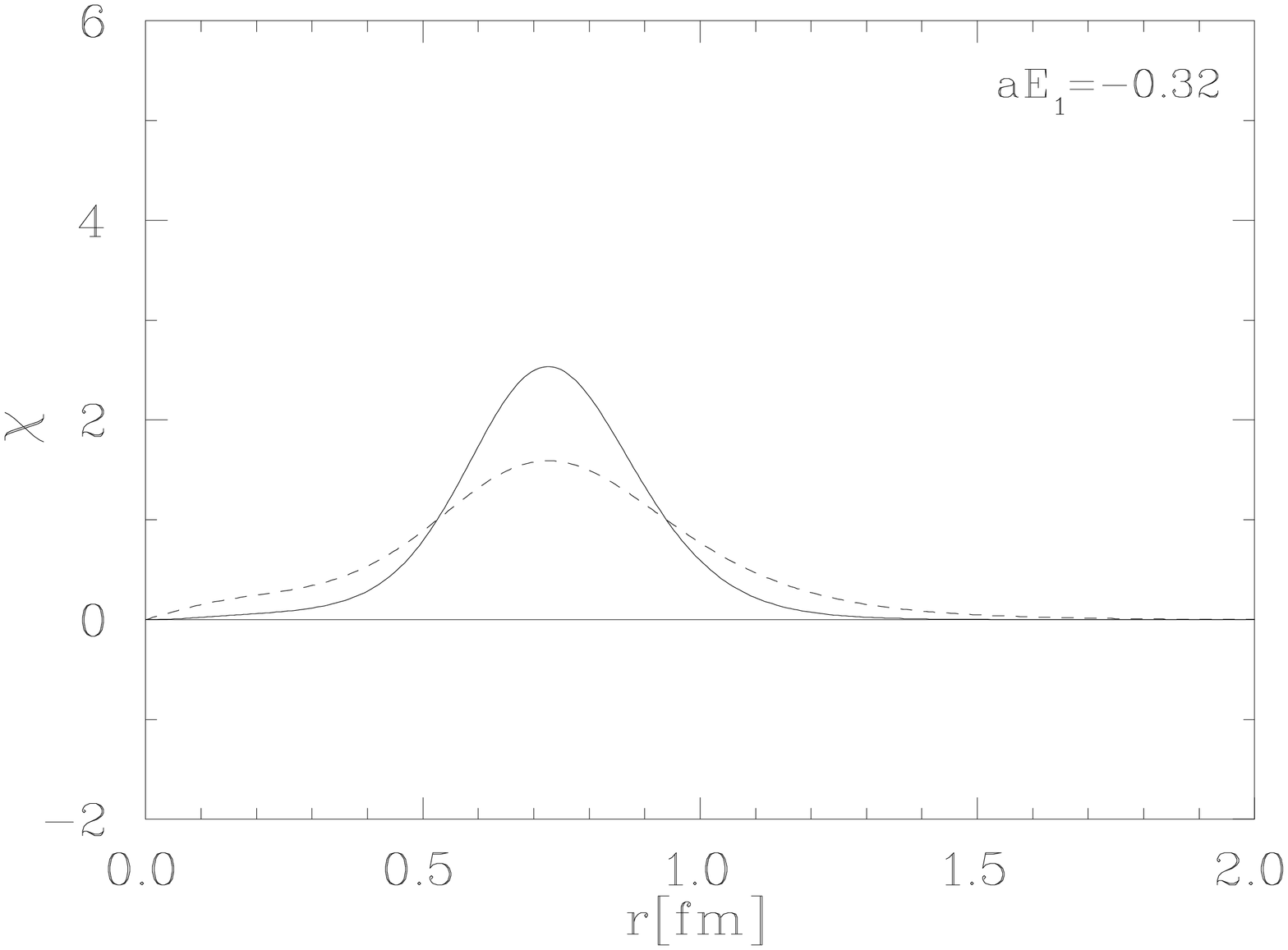}\hspace{4mm}
\caption{S-wave bound state wave functions $\chi=u_a(r)$ (dashed lines) and their squares
$\chi=|u_a(r)|^2$ (solid lines) for the $D$--$\Lambda_c$ system.}
\label{fig:5c}\end{figure}

\begin{figure}
\center
\includegraphics[angle=90,width=72mm]{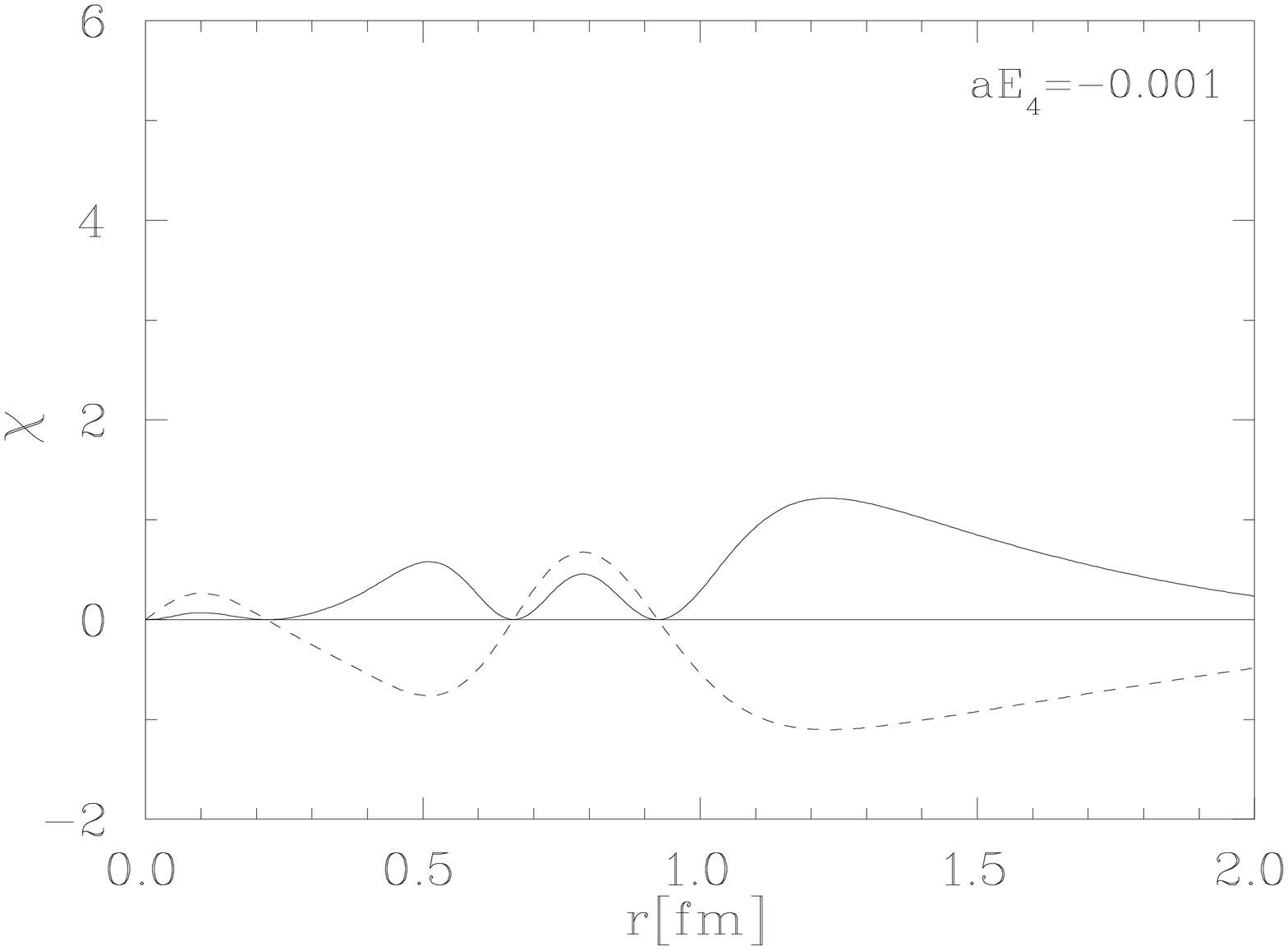}\hspace{4mm}\vspace{2.8mm}\\
\includegraphics[angle=90,width=72mm]{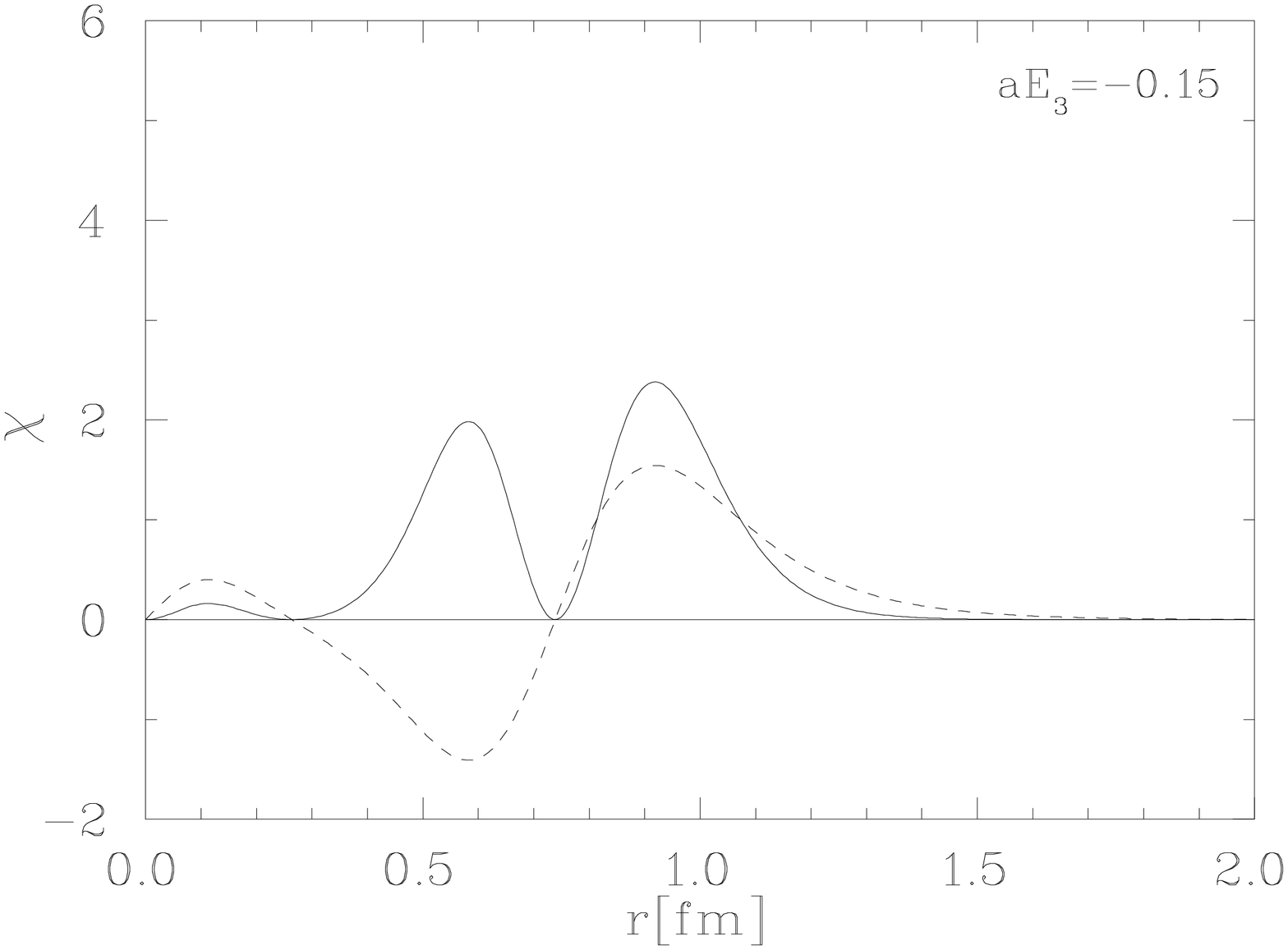}\hspace{4mm}\vspace{2.8mm}\\
\includegraphics[angle=90,width=72mm]{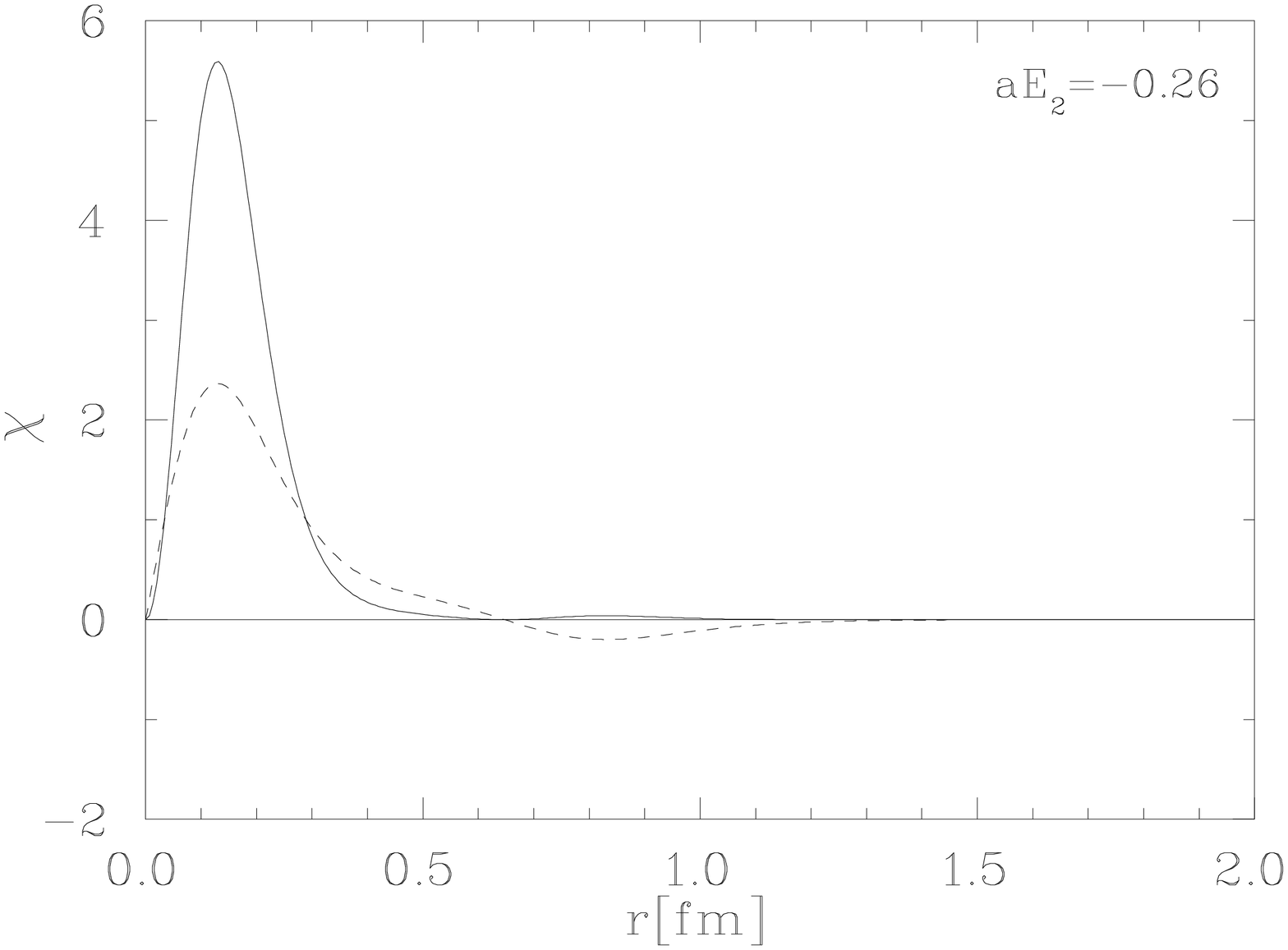}\hspace{4mm}\vspace{2.8mm}\\
\includegraphics[angle=90,width=72mm]{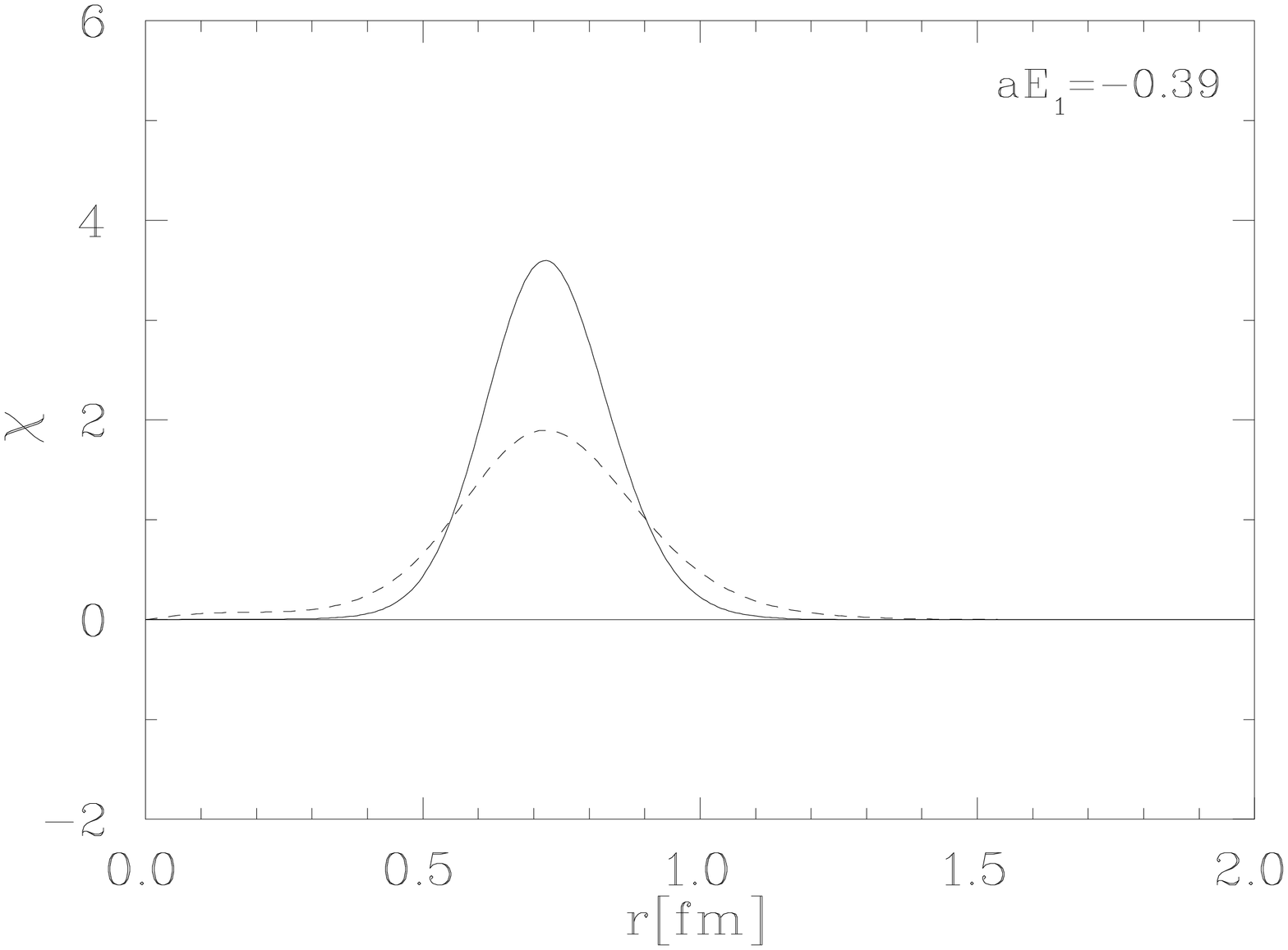}\hspace{4mm}
\caption{S-wave bound state wave functions $\chi=u_a(r)$ (dashed lines) and their squares
$\chi=|u_a(r)|^2$ (solid   lines) for the $B$--$\Lambda_b$ system.}
\label{fig:5b}\end{figure}

The S-wave bound state wave functions $u_a(r)$ and their squares $|u_a(r)|^2$ are displayed
in Figs.~\ref{fig:5s},\ref{fig:5c} and \ref{fig:5b}.
The wave function of the $K$--$\Lambda$ system, Fig.~\ref{fig:5s}, peaks at around
$r\sim 0.6$--$0.9$fm. This is a typical nuclear physics distance scale, and clearly points
to the structure of a hadronic molecule.
With increasing (heavy) quark mass the number of bound states also increases.
The ground state of the $D$--$\Lambda_c$ system, Fig.~\ref{fig:5c}, exhibits the same
feature. However, the excited state wave function has sizable support at small
distances $r\lesssim 0.2$fm. The wave function starts to `feel' the short-range attraction
of the potential $V_a(r)$, see Fig.~\ref{fig:Vpot}(a). 
This property is fully developed in $B$--$\Lambda_b$, see Fig.~\ref{fig:5b}.
The ground state wave function `lives' in the large-distance trough of $V_a(r)$ and
thus points to a hadronic molecule. On the other hand, the wave function of first
excited state `lives' in the small-distance regime of $V_a(r)$, which indicates a very
tight spacial structure. There is nothing as distinct about the third state, and the forth
one strongly resembles a continuum state.

The wave functions of the ground and first excited $B$--$\Lambda_b$ states
with $aE_1=-0.39$ and $aE_2=-0.26$, respectively, shown
in Fig.~\ref{fig:5b} deserve further comment.
Viewing the system as consisting of five quarks, the ground state is clearly a
pure two-hadron molecule, and the rms-radius evaluates to $r_{\rm rms}\approx 0.74$fm.
The excited state, on the other hand, is spatially compact
in a dramatic way: Most of the probability density $|u_a(r)|^2$ is within a
region of less than $0.3$fm radius and peaks just above $0.1$fm.
Inevitably, this five-quark state should be interpreted as a pentaquark.

We should remark here that our original five-quark operators (\ref{Op2}) are indeed
capable of describing such physics, because the set comprises operators where the
heavy quark and anti-quark have very small distances, the smallest one
being the lattice constant $a=0.088$fm.
Also, the $B$--$\Lambda_b$ results are probably our most reliable ones, because
relativistic effects are not very important.

On the other hand, systematic errors, having to do with the adiabatic approximation,
extrapolations to zero pion mass, and dismissal of relativistic effects, clearly render
our interpretation a qualitative one.
It is interesting to ask though if the reason for
our inability to establish a clear experimental signal for a pentaquark
\cite{Hicks:2007vg,DeVita:2006ha}, and
also to irrefutably establish a lattice QCD pentaquark
\cite{Csikor:2005xb,Sasaki:2005kg,Mathur:2004jr}, could be
rooted in the assumption that the pentaquark shall be a ground state of
bound light quarks. There is no a priori reason for this. The well-publicized
quark model decouplet $\Theta$~particle \cite{Diakonov:2003jj} may well
have a molecule-like structure.

Thus we speculate that the pentaquark, if we understand it as a tight
five-quark bound state, may not exist for all light flavored (u,d,s) quarks,
but there is still hope in the heavy-flavor sector. However, a pentaquark may then
reveal itself only as an excited state, the ground state being a hadronic molecule. 

\begin{figure}
\center
\includegraphics[angle=90,width=72mm]{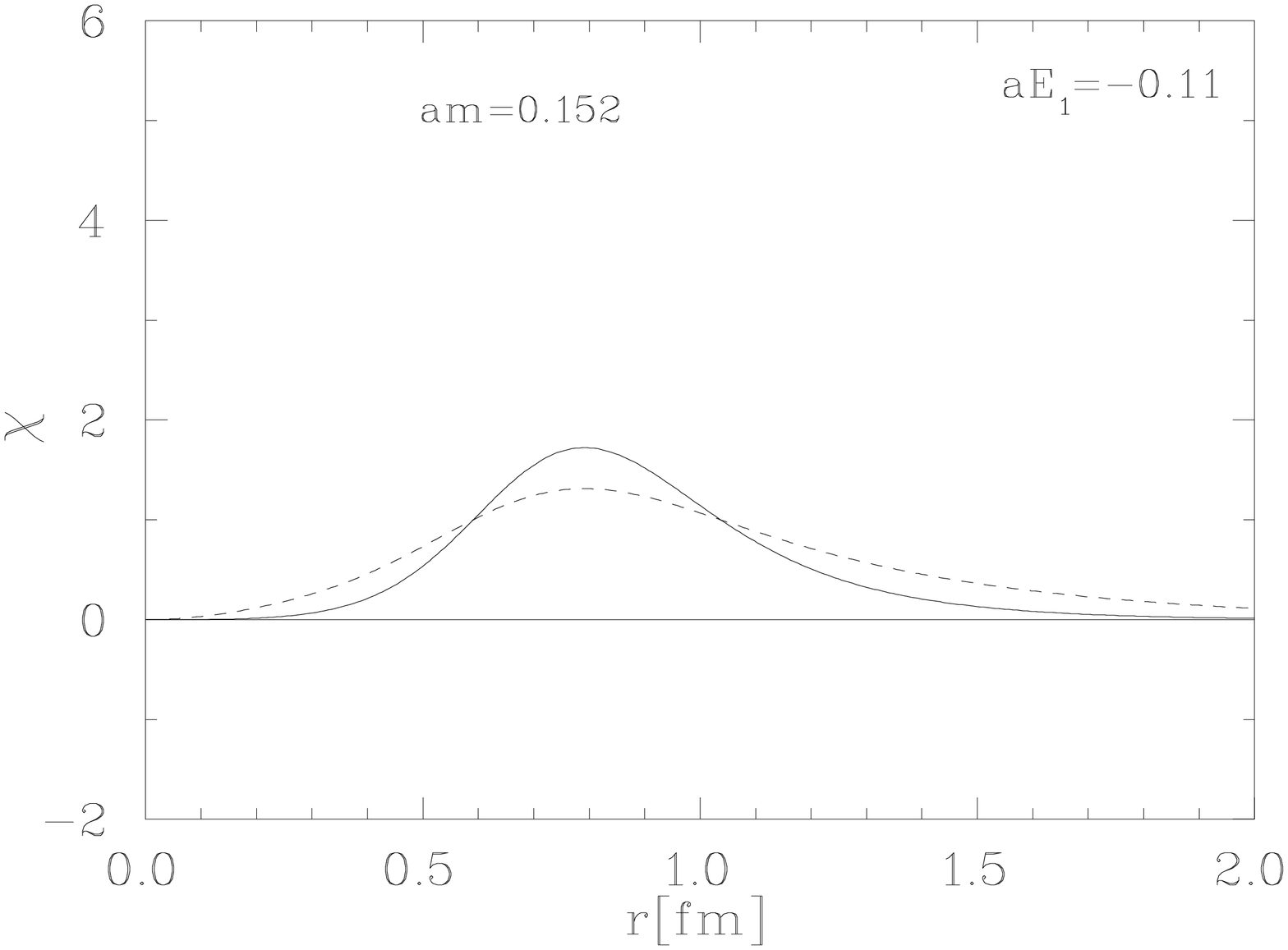}\hspace{4mm}\vspace{2.8mm}\\
\includegraphics[angle=90,width=72mm]{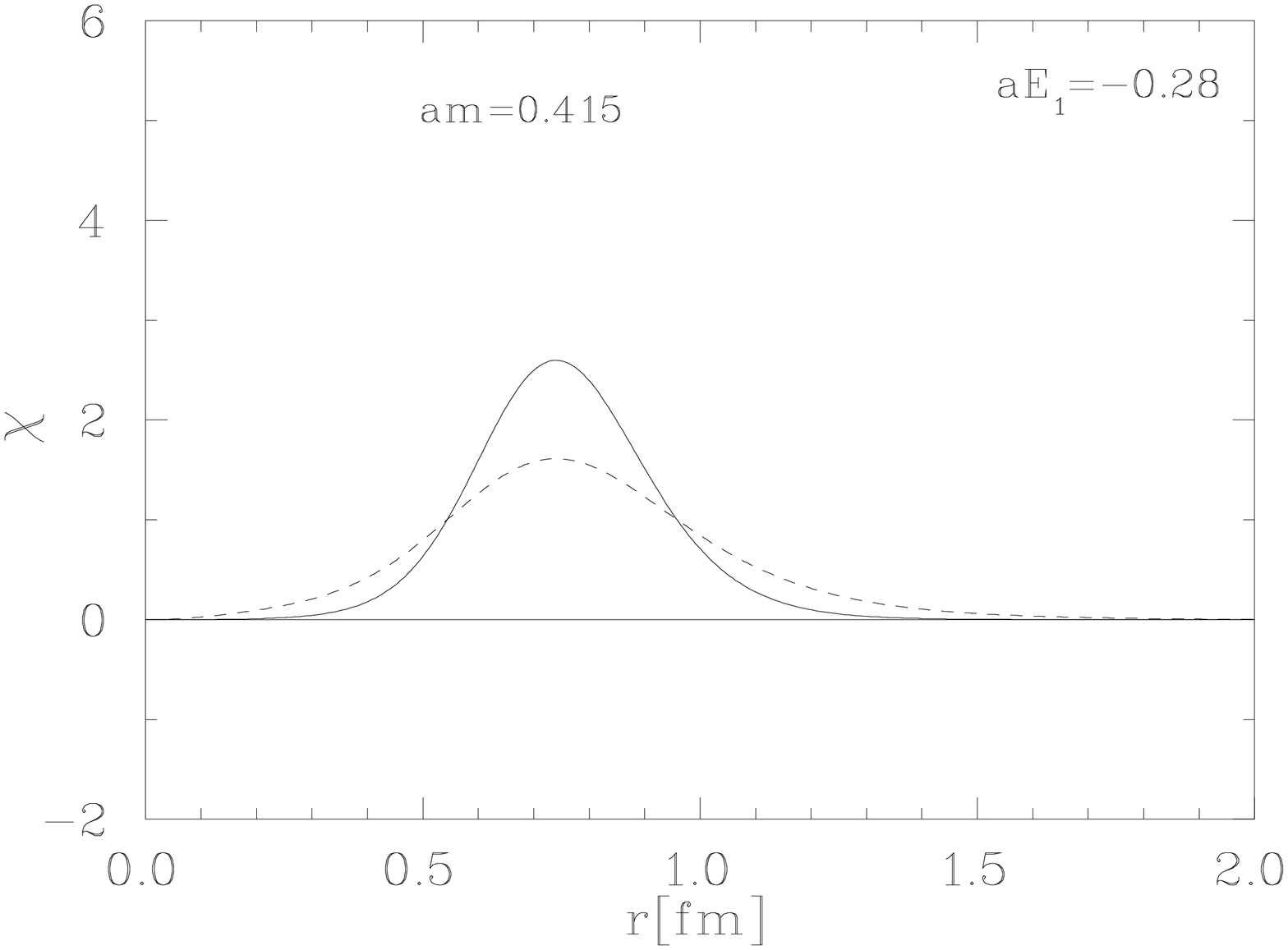}\hspace{4mm}\vspace{2.8mm}\\
\includegraphics[angle=90,width=72mm]{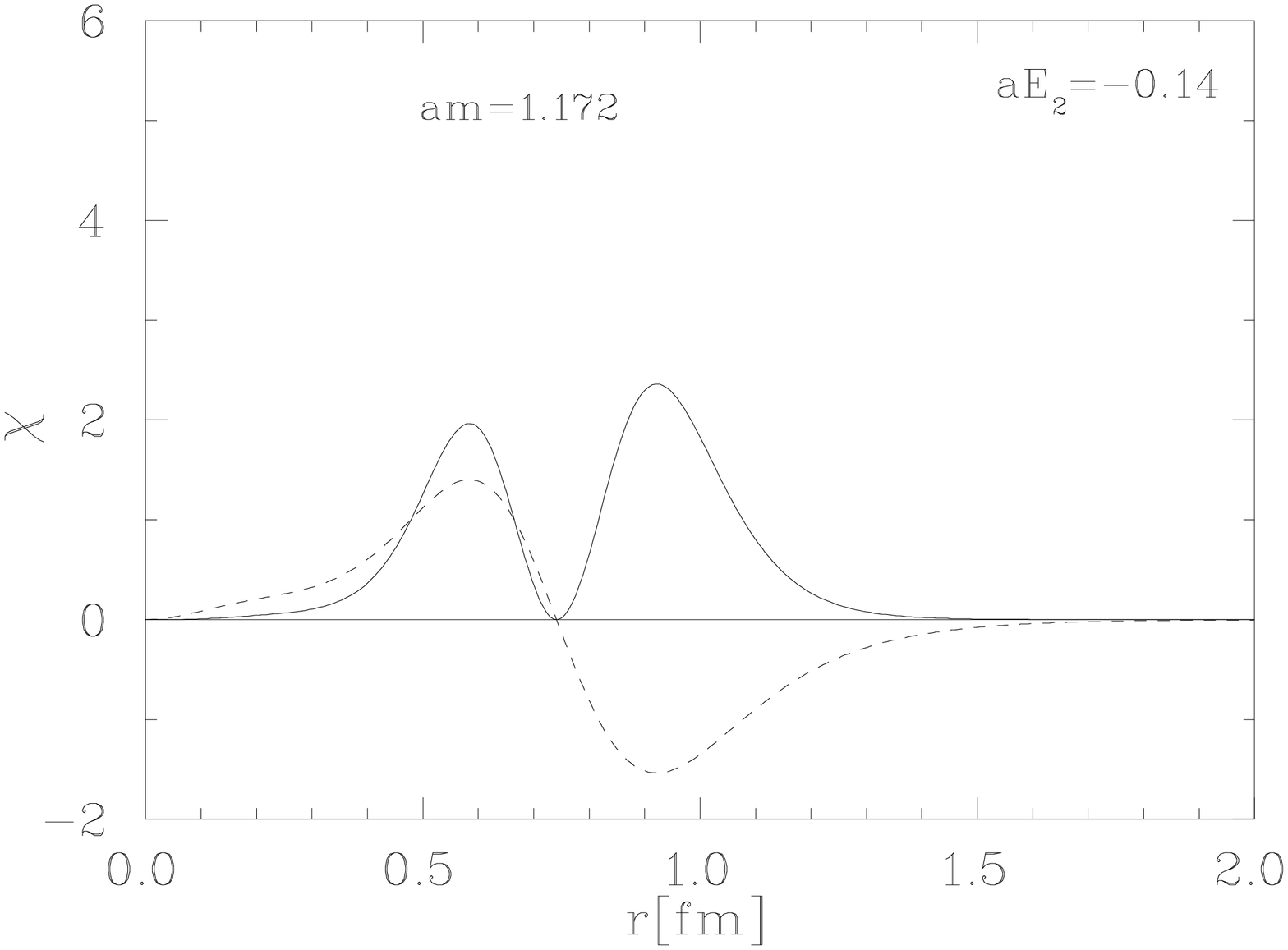}\hspace{4mm}\vspace{2.8mm}\\
\includegraphics[angle=90,width=72mm]{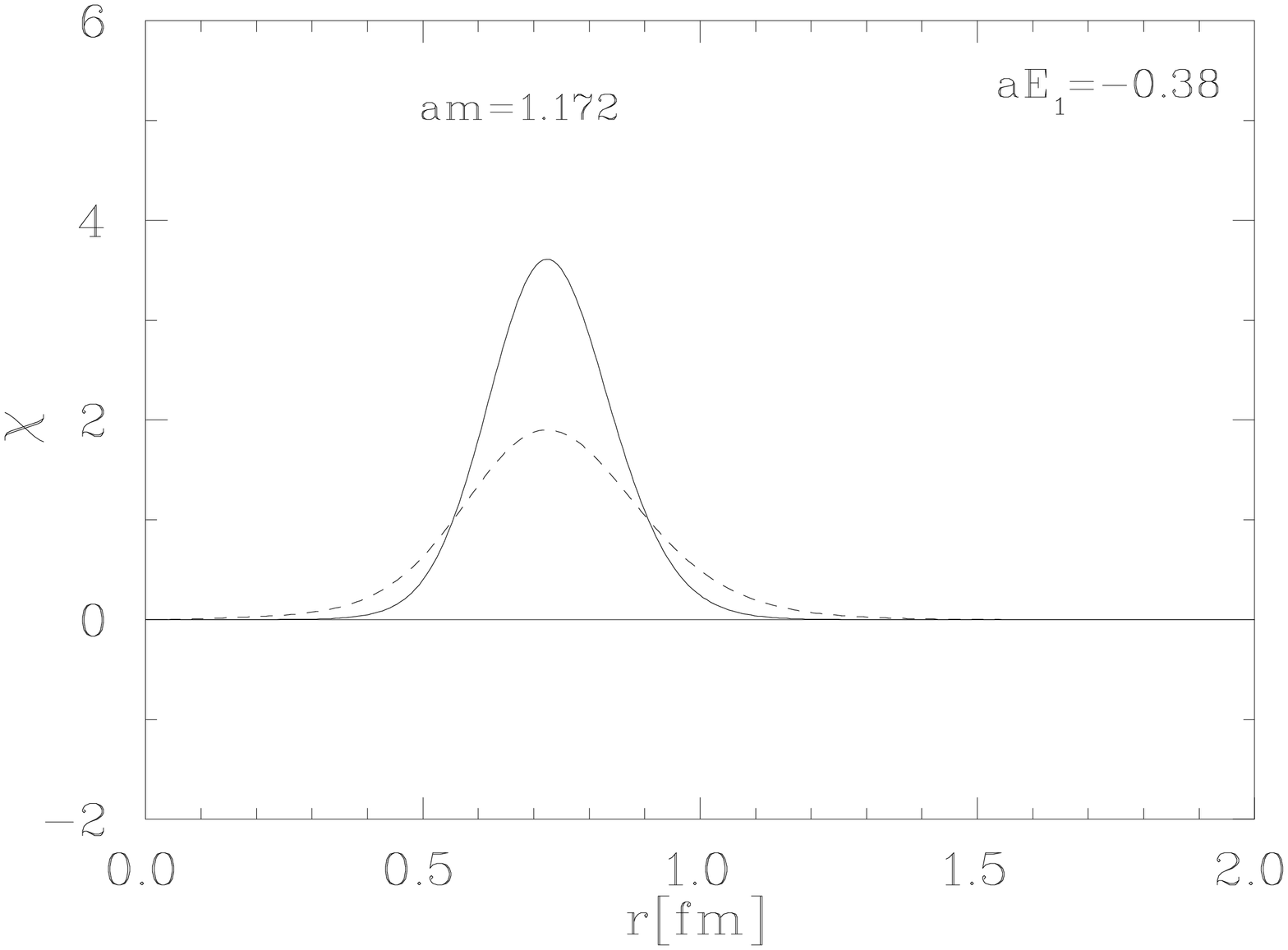}\hspace{4mm}
\caption{P-wave bound state wave functions $\chi=u_a(r)$ (dashed lines) and their squares
$\chi=|u_a(r)|^2$ (solid   lines).
The insets for $am$ and $aE_n$ match the entries of Tab.~\protect\ref{tab:bstate}.}
\label{fig:6P}\end{figure}

\begin{figure}
\center
\includegraphics[angle=90,width=72mm]{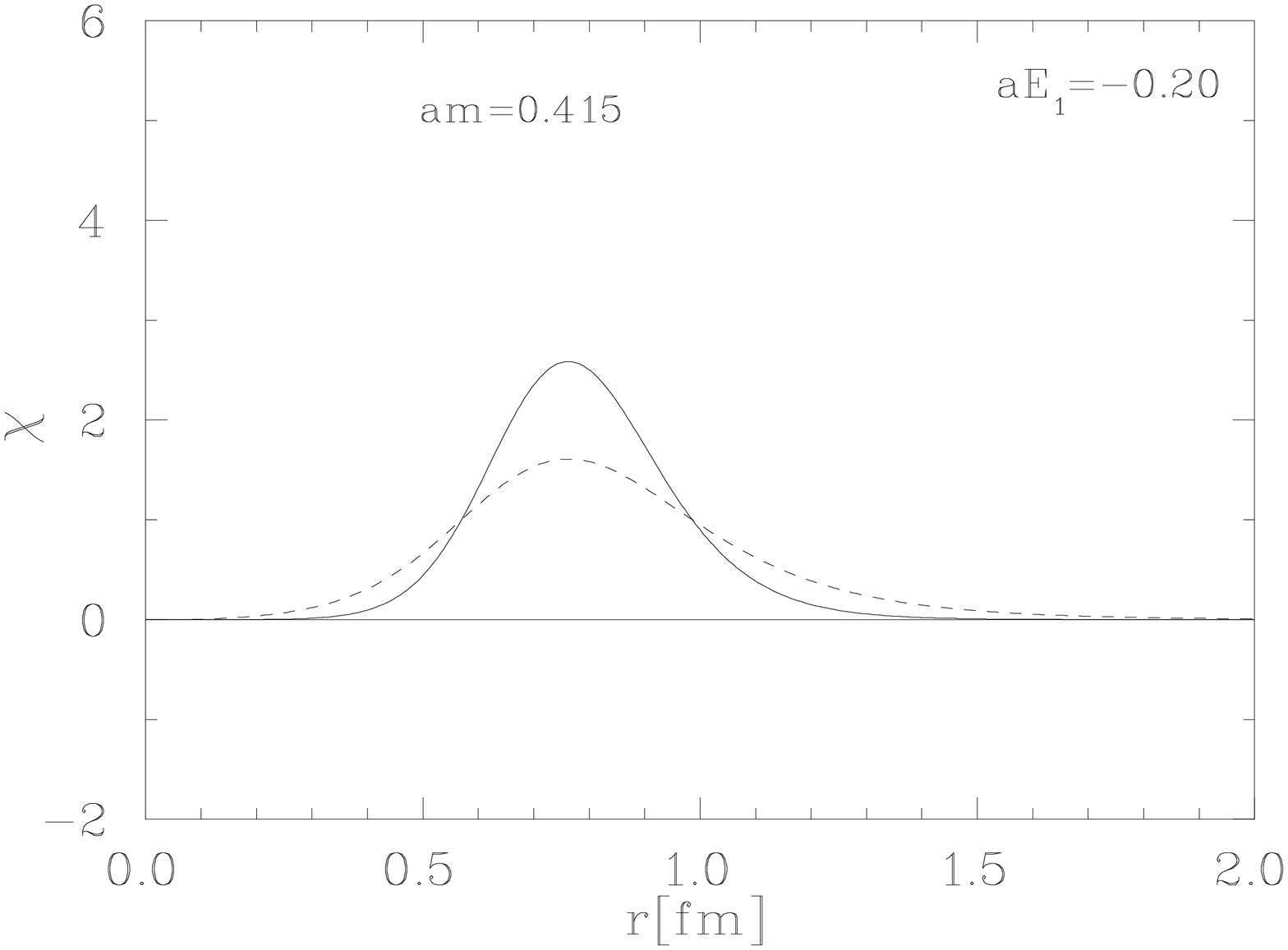}\hspace{4mm}\vspace{2.8mm}\\
\includegraphics[angle=90,width=72mm]{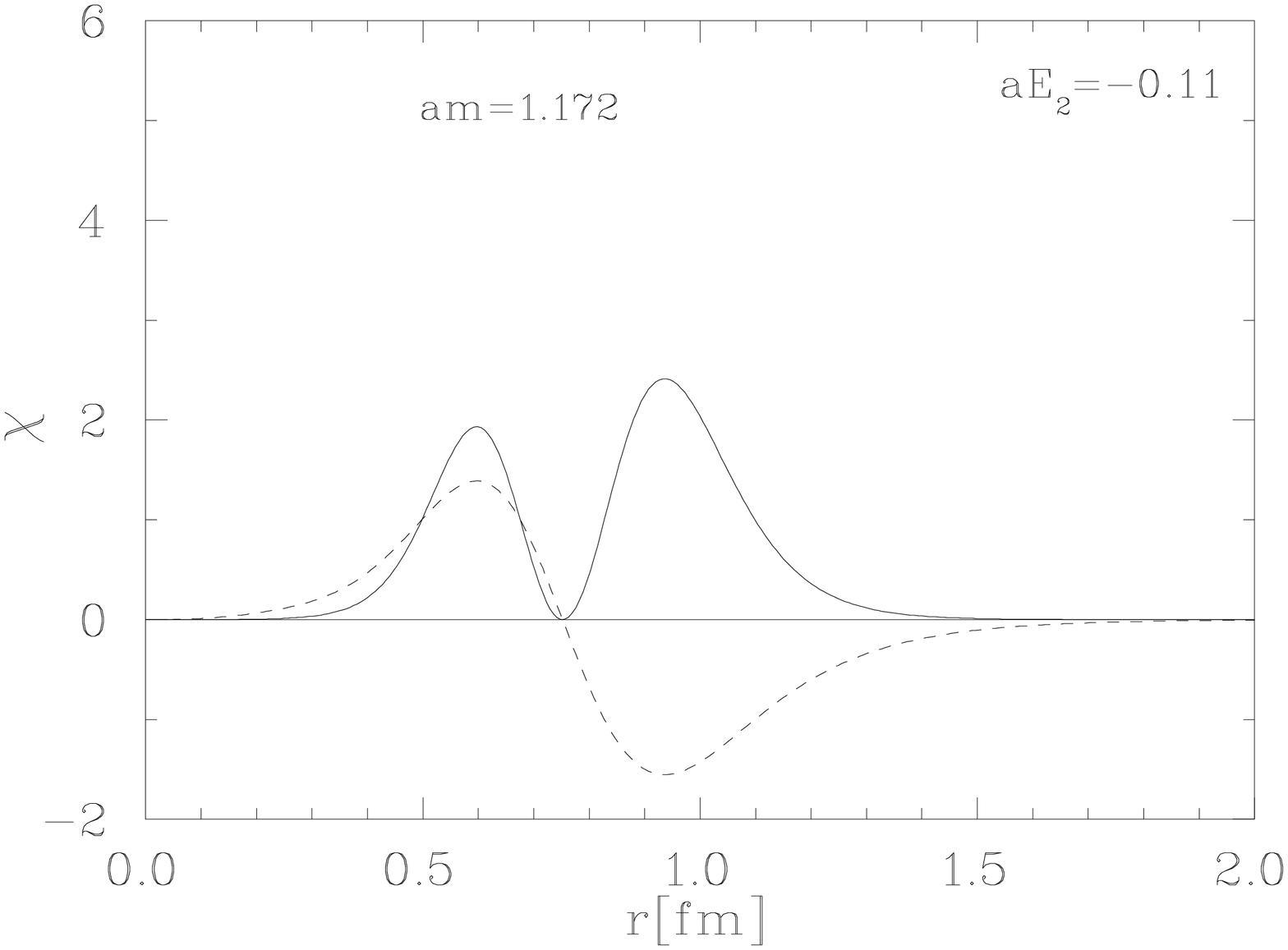}\hspace{4mm}\vspace{2.8mm}\\
\includegraphics[angle=90,width=72mm]{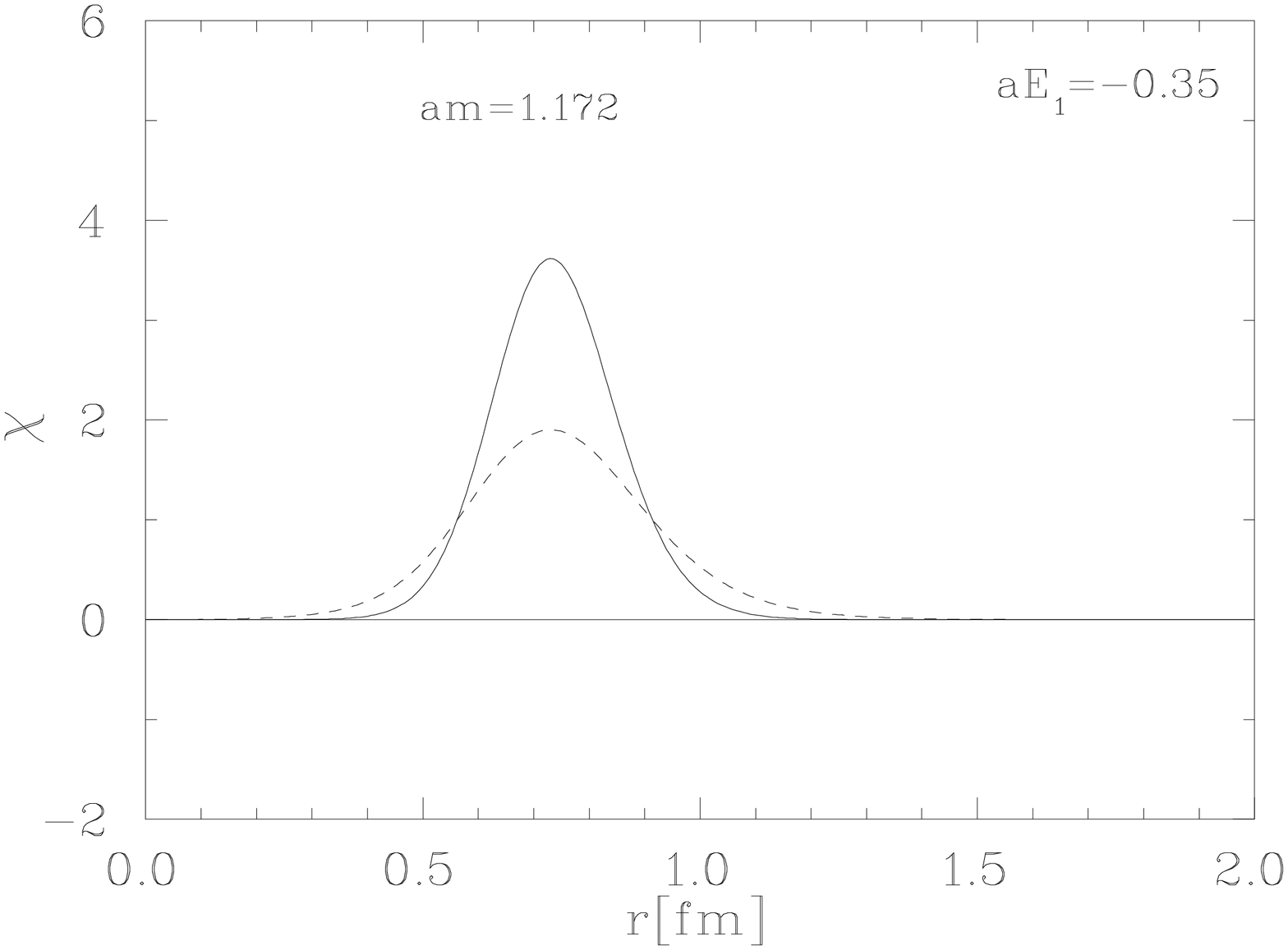}\hspace{4mm}
\caption{D-wave bound state wave functions $\chi=u_a(r)$ (dashed lines) and their squares
$\chi=|u_a(r)|^2$ (solid   lines).
The insets for $am$ and $aE_n$ match the entries of Tab.~\protect\ref{tab:bstate}.}
\label{fig:6D}\end{figure}

Finally, as a matter of course, we show in Figs.~\ref{fig:6P} and \ref{fig:6D} the
$P$- and $D$-state wave functions of the calculation.
And in Tab.~\ref{tab:bstatemasses} we have compiled physical mass spectra of the
considered five-quark systems by subtracting the computed binding energies
from the experimental single meson and baryon masses \cite{Yao:2006px}.
We give those numbers for completeness only, without claiming quantitative relevance. 
\begin{table}
\caption{\label{tab:bstatemasses}Physical five-quark hadron masses $m_n$, in GeV,
for three values of the reduced mass $m$ corresponding to the systems
$K$--$\Lambda$, $D$--$\Lambda_c$, and $B$--$\Lambda_b$. The layout matches that
of Tab.~\protect\ref{tab:bstate}.}
\begin{ruledtabular}
\begin{tabular} {cccccc}
$\ell$ & $m$ & $m_1$ & $m_2$ & $m_3$ & $m_4$ \\ \hline
0 & 0.342 &  1.11(28) &           &           &           \\
  & 0.934 &  3.44(29) &  4.07(29) &           &           \\
  & 2.639 & 10.02(30) & 10.33(35) & 10.58(27) & 10.90(20) \\ \hline
1 & 0.342 &  1.36(26) &           &           &           \\
  & 0.934 &  3.54(29) &           &           &           \\
  & 2.639 & 10.05(30) & 10.60(27) &           &           \\ \hline
2 & 0.342 &           &           &           &           \\
  & 0.934 &  3.70(29) &           &           &           \\
  & 2.639 & 11.11(30) & 10.66(26) &           &
\end{tabular}
\end{ruledtabular}
\end{table}

\section{Summary and conclusion}

We have performed a lattice simulation of a five-quark $K$--$\Lambda$ like hadronic system
on an anisotropic and asymmetric lattice. The heavy quark anti-quark pair was treated as
static. Thus it was possible to compute the total energy of the system as a function
of the relative distance $r$ between the hadrons. 
The maximum entropy method was employed toward this end.
An extrapolation into the physical pion mass region was also performed.
The objective was to extract an adiabatic potential $V_a(r)$ for the relative motion.
The potential turns out to have two distinct attractive troughs, one at intermediate distances,
$r\approx 0.7$fm, and one at short range, $r\lesssim 0.2$fm.

To study the dynamics of the two-hadron system we have used $V_a(r)$ in a
Schr{\"o}dinger equation for three values of the physical reduced mass,
corresponding to $K$--$\Lambda$, $D$--$\Lambda_c$, and $B$--$\Lambda_b$.
With increasing heavy-quark mass $m_s<m_c<m_b$ the number of S-wave bound
states increases from one to four. We have examined the corresponding
wave functions to study the nature of the five-quark systems.

Systematic errors predominantly originating from using the adiabatic
approximation, extrapolation to zero pion mass, the non-relativistic
framework, and possibly the quenched approximation, render the results
of our study qualitative only.
Nevertheless, we believe that the results for the $B$--$\Lambda_b$ ground
and excited state wave functions, respectively, bring to light a particularly
interesting scenario with regard to five-quark system physics: 
The ground state is best described as a hadronic molecule, with a relative
hadron-hadron distance matching a nuclear physics scale, whereas the excited
state exhibits a wave function with support on very short distances of
$r\lesssim 0.2$fm, or so. This leads us to interpret the excited state as
a pentaquark. The results for the $b$-quark system are less prone to
relativistic corrections and therefore are our most reliable ones.
The light-quark $K$--$\Lambda$ exhibits one ground state which clearly is a
hadronic molecule.

In the light of our results it might be worthwhile to initiate future
lattice QCD studies with this scenario in mind, but with a more realistic
set of operators, lattices, actions, and lighter quark masses, etc.
All things considered, there is ground for the hypothesis that an all light-quark
pentaquark may not exist, but that the existence of a genuine
pentaquark in the heavy-quark sector cannot be ruled out. However, if
present, it may well be an excited state. 

\begin{acknowledgments}
This material is based upon work supported by the National Science Foundation
under Grant No. 0300065.
\end{acknowledgments}

\end{document}